\documentclass[aps,prc,twocolumn,superscriptaddress,nofootinbib,floatfix]{revtex4-1}
\usepackage{enumitem}
\usepackage{amsmath,amssymb}
\usepackage[utf8]{inputenc}
\usepackage{graphicx}
\usepackage{bm}
\usepackage{hyperref}
\hypersetup{colorlinks=true,linkcolor=blue,citecolor=blue,urlcolor=blue}

\newcommand{\lr}[1]{\left\langle #1\right\rangle}

\newcommand{\pT}{p_{\mathrm{T}}}
\newcommand{\dpt}{\delta p_{\mathrm{T}}}
\newcommand{\HIJING}{\textsc{hijing}}
\newcommand{\PYTHIA}{\textsc{pythia}}
\newcommand{\obs}{\mathrm{obs}}
\newcommand{\nf}{\mathrm{nf}}

\newcommand{\LM}{\mathrm{LM}}
\newcommand{\pp}{\mathrm{pp}}
\newcommand{\Ocal}{\mathcal{O}}
\newcommand{\Ecal}{\mathcal{E}}
\newcommand{\snn}{\sqrt{s_{\mathrm{NN}}}}
\newcommand{\OO}{\mathrm{O}\!+\!\mathrm{O}}
\newcommand{\dAu}{d\!+\!\mathrm{Au}}

\newcommand{\pPb}{p\!+\!\mathrm{Pb}}

\graphicspath{{./}{fig/}}

\begin{document}

\title{Nonflow Subtraction Beyond Two-Particle Correlations}
\newcommand{\moe}{Key Laboratory of Nuclear Physics and Ion-beam Application (MOE), and Institute of Modern Physics, Fudan University, Shanghai 200433, China} \newcommand{\fudan}{Shanghai Research Center for Theoretical Nuclear Physics, NSFC and Fudan University, Shanghai 200438, China} \newcommand{\fudanP}{Physics Department and Center for Particle Physics and Field Theory, Fudan University, Shanghai 200438, China}
\newcommand{\sbu}{Department of Chemistry, Stony Brook University, Stony Brook, NY 11794, USA}\newcommand{\bnl}{Physics Department, Brookhaven National Laboratory, Upton, NY 11973, USA}

\author{Zaining Wang}\affiliation{\moe}\affiliation{\sbu}
\author{Jiangyong Jia}\email{jiangyong.jia@stonybrook.edu}\affiliation{\sbu}\affiliation{\bnl}
\author{Jinhui Chen}\email{chenjinhui@fudan.edu.cn}\affiliation{\moe}\affiliation{\fudan}
\author{Shengli Huang}\email{shengli.huang@stonybrook.edu}\affiliation{\sbu}
\author{Chunjian Zhang}\email{chunjianzhang@fudan.edu.cn}\affiliation{\moe}\affiliation{\fudan}
\author{Zhengxi Yan}\affiliation{\sbu}
\date{\today}

\begin{abstract}
Establishing collective flow in small collision systems is a crucial and necessary condition for pinning down the minimum conditions for quark-gluon plasma (QGP) formation. In two-particle correlations, nonflow has been subtracted with good control, pushing the reach of flow measurements down to very small event particle multiplicity $N$. However, the multi-particle nature of collectivity has not been established in the same $N$ regime, because the residual nonflow surviving the subevent procedure in multi-particle cumulants has never been quantified.  We develop a general nonflow subtraction framework for $m$-particle cumulants, built around the approximate $1/N^{m-1}$ scaling of nonflow in the independent-source picture. Correlators containing $v_1$ serve as nonflow-dominated estimators, since the $p_{\mathrm T}$-integrated dipolar flow is expected to be small. Using \HIJING{} as a controlled nonflow-only environment, we test the subtraction for three target observables ($\langle v_2^2\rangle$, $\langle v_2^2\delta p_{\mathrm T}\rangle$, and $c_2\{4\}$) in O+O at $\sqrt{s_{\mathrm{NN}}} = 5.36$~TeV and 200~GeV and in $d$+Au at $\sqrt{s_{\mathrm{NN}}} = 200$~GeV. Most of the nonflow is removed, with residual fractions typically within 20--30\% when converted to the two-particle level, though the best estimator differs across the three targets. We identify a multiplicity-reweighting correction, previously overlooked in two-particle correlations, that explains the long-standing undersubtraction of the naive $1/N$-scaling method; its impact grows as a power of the correlator order. The framework gives a systematic route to nonflow subtraction beyond two-particle correlations, broadening the class of multi-particle observables accessible to the small-system flow program.
\end{abstract}

\maketitle
\tableofcontents
\numberwithin{equation}{section}

%==============================================================
\section{Introduction}
\label{sec:intro}
The observation of long-range azimuthal correlations in high-energy $pp$, $p$+A, and O+O collisions has firmly established the presence of collective flow in small systems~\cite{CMS:2010ifv,CMS:2012qk,ATLAS:2012cix,ALICE:2012eyl,PHENIX:2013ktj,ATLAS:2014qaj,CMS:2015yux, ATLAS:2017rtr,ATLAS:2019vcm,ATLAS:2021jhn, ALICE:2024vzv}, and has been used to argue for QGP formation in these systems~\cite{Dusling:2015gta,Nagle:2018nvi,Busza:2018rrf,Schenke:2021mxx}.

In two-particle correlations (2PC), nonflow contributions arising from jet fragmentation, resonance decays, and Hanbury-Brown-Twiss (HBT) effects can be suppressed by requiring a pseudorapidity gap between the particles~\cite{ATLAS:2014qaj,ATLAS:2015hzw,CMS:2013jlh}. In the independent-source picture, each source correlates only a few particles while the number of sources grows with the event particle multiplicity $N$ in the analyzed acceptance. Nonflow is therefore diluted approximately as $1/N$, and its relative importance grows towards smaller systems and lower $N$~\cite{Borghini:2001vi}. The remaining nonflow can then be estimated and subtracted using minimum-bias $pp$ or low-multiplicity collisions as a reference~\cite{ATLAS:2014qaj,ATLAS:2015hzw,STAR:2022pfn}. A scaled reference subtraction removes most of the nonflow in high-multiplicity events. The residual biases arising from multiplicity selection can be further constrained by employing different scaling prescriptions, including the multiplicity scaling (or $c_0$ method), the dipole scaling method (or $c_1$ method), the near-side jet scaling, and the template method~\cite{ATLAS:2014qaj,ATLAS:2015hzw,CMS:2013jlh,ATLAS:2018ngv}, whose spread brackets the nonflow systematic uncertainty. These techniques have established significant $v_2$--$v_4$ signals down to very small $N$~\cite{ATLAS:2018ngv,STAR:2022pfn,PHENIX:2017nvb}.

However, an analogous framework for multi-particle cumulants (MPC) is lacking. In MPC the prevailing practice relies on the so-called subevent method~\cite{Jia:2017hbm} to suppress nonflow without a subsequent explicit subtraction. This approach has important limitations. 

First, without an explicit estimate of the residual nonflow, one cannot quantify the remaining systematic uncertainty in a given measurement, nor distinguish residual nonflow from flow decorrelation in $\pT$ and $\eta$~\cite{Bozek:2015bna,Jia:2014vja,Mazeliauskas:2015vea}, nor extend measurements to low $N$ where nonflow dominates even after subevent suppression. Measurements of single-harmonic multi-particle cumulants~\cite{ATLAS:2017rtr,CMS:2019lin,ALICE:2019zfl}, symmetric/asymmetric cumulants~\cite{ALICE:2019zfl,ATLAS:2018ngv,CMS:2019lin,CMS:2017kcs, ATLAS:2017rtr}, and the anisotropic--radial flow correlators~\cite{ATLAS:2022dov,CMS:2024rvk,ALICE:2026cly} in small systems all exhibit strong sensitivity to subevent choice and event-class definition that is qualitatively consistent with unsubtracted nonflow.

Second, the subevent method suppresses nonflow at the cost of statistical power. The four-subevent four-particle cumulant loses a factor of $4^4/4! \approx 11$ in the number of unique quadruplets~\cite{Jia:2017hbm}; an explicit subtraction would allow the higher-statistics standard (full-acceptance) or two-subevent method to be used while still controlling nonflow.

The multi-particle nature of collective flow is its defining feature: genuine collectivity should produce significant, mutually consistent signals across all multi-particle cumulants~\cite{Borghini:2001vi,Bilandzic:2010jr,Bilandzic:2013kga,Bozek:2016yoj,ATLAS:2022dov}. Reliably quantifying and subtracting nonflow in these observables is needed to access the detailed properties of collectivity, and its response to initial geometry and nuclear structure, in small systems~\cite{STAR:2022pfn, Welsh:2016siu,Schenke:2020mbo,Zhao:2022ugy, Giacalone:2021udy,Zhang:2024vkh}. With rich datasets covering many energies and systems now available at RHIC and the LHC, and with O+O collisions recently collected at both facilities~\cite{ATLAS:2021jhn,ALICE:2024vzv,ATLAS:2025nnt,ALICE:2025luc,CMS:2025tga,STAR:2025ivi,Chen:2026gka}, extending nonflow subtraction beyond 2PC is timely and necessary.

Our approach extends the philosophy of 2PC nonflow subtraction to MPC. The guiding principle is the independent-source scaling of nonflow, $\Ocal\{m\}^{\nf} \sim 1/N^{m-1}$, for an $m$-particle cumulant $\Ocal\{m\}$. Cumulants containing $v_1$ are natural nonflow estimators on two counts. First, in the long-range domain, after the rapidity gap removes the near-side jet, the surviving low-$\pT$ nonflow from jets, decays, and HBT projects onto a dipolar shape via global momentum conservation (GMC), and the resulting multi-particle correlator scales as $\sim 1/N^{m-1}$~\cite{Borghini:2003ur}. Second, the genuine $\pT$-integrated dipolar flow is small, because the intrinsic dipolar flow changes sign around $\pT \approx \lr{\pT}$ and integrates to nearly zero~\cite{Borghini:2002mv,Luzum:2010fb,Teaney:2010vd,ATLAS:2014qaj,Jia:2012gu}. 

The philosophy of using a known dominant-nonflow observable to estimate the scaling of nonflow in the target observable underpins the 2PC subtraction~\cite{Lim:2019cys,Grosse-Oetringhaus:2024bwr} and generalizes naturally to higher orders. This line of reasoning provides a family of physically motivated nonflow scaling estimators.

Our numerical study uses the \HIJING{} event generator~\cite{Gyulassy:1994ew}, which contains few-particle correlations from minijet production, jet and string fragmentation, and resonance decays but no genuine collective flow. It has previously been used as a nonflow-only reference, primarily to study qualitative trends in two-particle correlations~\cite{Ollitrault:2009ie,Lim:2019cys,STAR:2023wmd,Feng:2024eos}. The aim is not to use \HIJING{} to predict the absolute nonflow in data: its particle multiplicity is too high by roughly 50\%~\cite{ALICE:2021hkc,ALICE:2025woy}, and its correlations have the wrong rapidity structure~\cite{STAR:2023wmd}. Instead, we use it as a controlled nonflow-only environment to test the practical implementation of the independent-source scaling and identify which reference estimator performs best. We propose that, under the assumption that it is transferable, the \emph{fractional} non-closure derived from \HIJING{}, rather than the absolute nonflow yield, can be propagated to data to provide a meaningful estimate of the subtraction uncertainty. This procedure is analogous to the use of \HIJING{} in the STAR 2PC analysis~\cite{STAR:2023wmd}. Nevertheless, the transfer assumption is model dependent and should be tested with other generators and data-driven closure comparisons.

We focus on the three target observables most relevant for small-system collectivity: $\lr{v_2^2}$, the two-particle elliptic flow; $\lr{v_2^2\,\dpt}$, the three-particle elliptic-flow--mean-$\pT$ covariance~\cite{Bozek:2016yoj,ATLAS:2022dov,Schenke:2020uqq}; and $c_2\{4\}$, the four-particle elliptic-flow cumulant. Results are presented for O+O at LHC energy ($\snn = 5.36$~TeV) and for both O+O and $d$+Au at RHIC energy ($\snn = 200$~GeV). We focus on the elliptic harmonic $n=2$ throughout; the extension to $v_3$- or $v_4$-related observables is deferred to future work.

The paper is organized as follows. Section~\ref{sec:2pc} reviews nonflow subtraction in 2PC. Section~\ref{sec:motivation} discusses the motivation for extending nonflow subtraction to MPC. Section~\ref{sec:framework} introduces the general scaling framework and the family of subtraction methods. Section~\ref{sec:analysis} describes the \HIJING{} analysis setup.  Section~\ref{sec:res} presents subtraction performance results and demonstrates the role of the multiplicity-reweighting factor. Section~\ref{sec:summary} concludes.
\section{Nonflow subtraction in two-particle correlations}
\label{sec:2pc}

The 2PC nonflow subtraction landscape consists of a family of related methods: the $c_0$-, $c_1$-, near-jet, and template variants, whose distinctions are subtle but consequential. Recent reviews~\cite{Lim:2019cys,Grosse-Oetringhaus:2024bwr,Feng:2024eos} provide useful surveys, but the relationships among the methods are at times blurred, and corrections that significantly affect the subtraction in practice, in particular the multiplicity-reweighting factor we identify in Sec.~\ref{sec:res3} and the impact of flow in the peripheral bin, are not sufficiently addressed. Because the multi-particle framework developed in Sec.~\ref{sec:framework} inherits the scaling assumptions and bias-correction structure of these 2PC methods directly, we devote this section to a uniform mini-review before generalizing to higher orders.

\subsection{Definitions and assumptions}

\emph{Flow}, in the simplest picture, refers to azimuthal correlations arising from the hydrodynamic response to the initial collision geometry~\cite{Heinz:2013th,Gale:2013da,Luzum:2013yya}, characterized by the harmonic coefficients $v_n$. In practice, several effects blur this simple picture. Subleading flow modes break the simple factorization assumed in the one-body description~\cite{Mazeliauskas:2015vea}. Jet wakes that thermalize in the medium contribute additional collective-like signals on top of the bulk flow~\cite{Yang:2022nei}. And at high $\pT$, path-length-dependent energy loss produces correlations with the collision geometry that have different response from bulk $v_n$~\cite{Gyulassy:2000gk}. These effects introduce both additional one-body and genuine many-body contributions to the measured harmonics.

\emph{Nonflow} refers to correlations unrelated to the collision geometry, arising from sources such as jet fragmentation, resonance decays, and HBT effects~\cite{Ollitrault:2009ie,Feng:2024eos}. All particle-production mechanisms obey GMC, which induces an event-wide dipolar correlation in addition to their own characteristic correlation structures. After the near-side jet is suppressed by the pseudorapidity gap, the surviving low-$\pT$ long-range nonflow is dominated by this GMC dipolar imprint. The separation between flow and nonflow is statistical and defined at the correlator level: a $\rho^0$ meson may carry collective flow and simultaneously induce nonflow correlations between its decay pions; a jet correlated with the impact parameter via path-length-dependent energy loss nevertheless produces multi-particle nonflow through its fragmentation. The flow and nonflow fractions are therefore observable-dependent and cannot be separated at the single-particle level~\cite{Mazeliauskas:2015vea,Mazeliauskas:2015efa}. The nonflow fraction must be defined and estimated separately for each $m$-particle correlator.

The subtraction framework further assumes that nonflow is \emph{unmodified} by the medium: resonance decays and jet fragmentation proceed as in vacuum. This assumption can fail when radial flow focuses the decay products of a boosted resonance, or when jet quenching induces diffusion wakes. Such medium modifications are a known limitation of the subtraction approach~\cite{STAR:2023wmd,Lim:2019cys} and contribute to the spread among different subtraction methods in data.

\emph{Notational convention.} Throughout this paper, $\lr{v_n^2}$ without a superscript denotes the genuine flow component, which is what the nonflow subtraction aims to recover. The observed quantity (long-range, after the $\Delta\eta$ gap is applied) is $\lr{v_n^2}^{\obs}$, and the nonflow component is $\lr{v_n^2}^{\nf}$, with the decomposition $\lr{v_n^2}^{\obs} = \lr{v_n^2}^{\nf} + \lr{v_n^2}$. In \HIJING{}, $\lr{v_n^2}=0$ and therefore $\lr{v_n^2}^{\obs} = \lr{v_n^2}^{\nf}$; any nonzero value returned by Eq.~\eqref{eq:2.1} is residual nonflow and quantifies the non-closure of the procedure.

\subsection{Subtraction formula and scaling}
\label{subsec:scaling}
In experimental analysis, a pseudorapidity gap is applied to suppress short-range correlations. The remaining long-range harmonics give the ``direct Fourier'' result $\lr{v_n^2}^{\obs}$. Nonflow is then subtracted:
\begin{equation}
  \lr{v_n^2}(N)  = \lr{v_n^2}^{\obs}(N)  - f(N) \times \lr{v_n^2}^{\obs}_{\scriptscriptstyle{\LM}},
  \label{eq:2.1}
\end{equation}
where $\LM$ denotes low-multiplicity events in the same system or minimum-bias $pp$ reference and $f$ is a method-dependent scaling factor that varies with $N$. In all experimental analyses to date, this scaling factor has been assumed common across all harmonics $n \geq 2$. We follow that convention here, noting that the underlying nonflow may in principle depend on the harmonic order.

In the independent-source picture~\cite{Borghini:2001vi}, nonflow scales as $1/N$, giving $f \sim N_\LM/N$. In small systems this scaling is distorted by multiplicity-selection bias: the steeply falling tail of $p(N)$ enriches high-$N$ events with extra-jet configurations, so nonflow at high $N$ exceeds the naive $1/N$ expectation and at low $N$ falls below it. The effect is stronger for near-side correlations, which need two fragments from the same jet, than for away-side correlations, which need only one particle from each side of the dijet~\cite{CMS:2013jlh,ATLAS:2015hzw}.

Equation~\eqref{eq:2.1} explicitly assumes that the $\LM$ events are devoid of genuine collective flow. This assumption is invalid: ATLAS and ALICE measurements, for instance, demonstrate that nonzero flow signals persist even in events with very low charged-particle multiplicity~\cite{ATLAS:2019vcm,ALICE:2023ulm}, and analogous evidence exists for peripheral $p$+A and A+A collisions. In response, experimental analyses developed methods that explicitly incorporate collectivity in the low-multiplicity bin, improving the treatment of nonflow subtraction and flow extraction.

\subsection{Methods that assume only nonflow in $\LM$}
There are three methods for estimating $f$ in Eq.~\eqref{eq:2.1}:
\begin{enumerate}
\item \textbf{Multiplicity- or $c_0$-scaling~\cite{ALICE:2012eyl,ATLAS:2012cix,PHENIX:2013ktj}:} 
\begin{align}
f_1 = N_{\LM}/N,
\label{eq:2.2}
\end{align}
assuming independent sources without selection bias.

\item \textbf{Dipolar- or $c_1$-scaling~\cite{PHENIX:2017nvb,STAR:2023wmd}:}
\begin{align}
f_2 = \lr{v_1^2}^{\obs}/\lr{v_1^2}^{\obs}_{\scriptscriptstyle{\LM}} \approx f_1 \times \frac{\text{Y}_\mathrm{away}}{\text{Y}_\mathrm{away,\scriptscriptstyle{LM}}},
\label{eq:2.3}
\end{align}
where $Y$ is the per-trigger yield integrated over a certain $\Delta\phi$ range around $\pi$. In the low-$\pT$ region, this effectively uses the ratio of dipolar flow between target and reference samples as a proxy for the scaling of the jet-induced contribution.

\item \textbf{Near-jet scaling~\cite{CMS:2013jlh,ATLAS:2014qaj,STAR:2015kak}:}
\begin{align}
f_3 = f_1 \times (\text{Y}_\mathrm{near}/\text{Y}_\mathrm{near,\scriptscriptstyle{LM}}),
\label{eq:2.4}
\end{align}
using the near-side jet per-trigger yield as the proxy~\cite{CMS:2013jlh}. This typically gives $f_3 > f_2$ due to the stronger multiplicity selection bias, which arises because the same jet must produce two fragments to form a near-side pair, as opposed to the $c_1$-scaling method, which requires only one particle of the pair to come from the away side.
\end{enumerate}

\subsection{Methods that account for flow in $\LM$}
\begin{enumerate}[start=4]
\item \textbf{Template method~\cite{ATLAS:2015hzw,STAR:2022pfn}:} 
The method starts with a Fourier decomposition of the per-trigger yield distribution
\begin{align}
\text{Y}(\Delta \phi)=\frac{N^{a}}{2\pi}(1+2 \sum_{n=1}^{\infty} \lr{v_n^2}^{\obs} \cos n \Delta \phi)  
\label{eq:2.5}
\end{align}
where $N^a \propto N$ is the integrated associated-particle yield per trigger. Nonflow is removed by subtracting the $\LM$ per-trigger yield distribution with a scale $F$, leaving an integrated residual pedestal $G$ modulated by $n \geq 2$ harmonics:
\begin{align}\label{eq:2.6a}
  \text{Y}(\Delta\phi)-F\,\text{Y}(\Delta\phi)_{\scriptscriptstyle{\LM}} = \frac{G}{2\pi}\left(1 + 2\sum_{n=2}^{\infty}\lr{v_n^2}^{\mathrm{tmp}}\cos n\Delta\phi\right).
\end{align}
The scale $F$ is fixed by requiring that the $n=1$ component cancel~\cite{Feng:2024eos},
\begin{align}\label{eq:2.6b}
N^a\lr{v_1^2}^{\obs}
  = F\,N^a_{\scriptscriptstyle{\LM}}\,
    \lr{v_1^2}^{\obs}_{\scriptscriptstyle{\LM}}
\Longrightarrow F = f_2/f_1\,,
\end{align}
which in turn fixes the pedestal,
\begin{align}\label{eq:2.6c}
G = N^a - F\,N^a_{\scriptscriptstyle{\LM}} = N^a\,(1-f_2)\,.
\end{align}

Because the pedestal is reduced, $G<N^a$, and all extracted harmonics are enhanced relative to the $c_1$-method. Inserting Eq.~\eqref{eq:2.5} into the left-hand side of Eq.~\eqref{eq:2.6a} makes this explicit:
\begin{align}\label{eq:2.7}
\lr{v_n^2}^{\mathrm{tmp}} =  \frac{\lr{v_n^2}^{\obs}
    - f_2\lr{v_n^2}^{\obs}_{\scriptscriptstyle{\LM}}}{1-f_2}\;.
\end{align}
The template method therefore differs from the $c_1$-method [Eq.~\eqref{eq:2.3}] by a factor of $1/(1-f_2)$. This rescaling recovers the true flow only when the genuine flow in the $\LM$ and target bins coincide, $\lr{v_n^2}_{\scriptscriptstyle{\LM}}=\lr{v_n^2}$, a condition that holds for $v_2$ in $pp$ collisions~\cite{ATLAS:2015hzw} but is violated in general. 

Historically this approach has been called the ``template-fit'' method, because the subtraction in Eq.~\eqref{eq:2.6a} was implemented through a $\chi^2$ minimization. As that procedure is mathematically equivalent to requiring the cancellation of the dipolar component, we drop the ``fit'' label for generality.

\item \textbf{Improved template method~\cite{ATLAS:2018ngv}:} 
When the flows in $\LM$ and the target bin differ, the template method is biased. The bias is easily seen by assuming nonflow is fully subtracted and setting the observed harmonics equal to the genuine flow ($\lr{v_n^2}^{\obs}\!\to\!\lr{v_n^2}$). Eq.~\eqref{eq:2.7} then becomes:
\begin{align}\label{eq:2.8}
\lr{v_n^2}^{\mathrm{tmp}} =  \frac{\lr{v_n^2}
    - f_2\lr{v_n^2}_{\scriptscriptstyle{\LM}}}{1-f_2}\;.
\end{align}

Clearly, this reproduces the true flow $\lr{v_n^2}$ if and only if $\lr{v_n^2}=\lr{v_n^2}_{\scriptscriptstyle{\LM}}$, which is the case for $v_2$ in $pp$ collisions. However, measurements have shown that all flow harmonics grow with multiplicity in $p$+A and A+A collisions, and that the triangular flow $v_3$ grows with multiplicity even in $pp$ collisions. In this case, the LM events have a smaller flow signal, $\lr{v_n^2}_{\scriptscriptstyle{\LM}}<\lr{v_n^2}$, and the template method systematically overestimates the signal, $\lr{v_n^2}^{\mathrm{tmp}}>\lr{v_n^2}$.

The improved template method removes this bias by rewriting Eq.~\eqref{eq:2.8}, expressing the true flow as a one-step correction to the template-method result:
\begin{align}\label{eq:2.9}
  \lr{v_n^2} = \lr{v_n^2}^{\mathrm{tmp}}- f_2
      \!\left(\lr{v_n^2}^{\mathrm{tmp}} -\lr{v_n^2}_{\scriptscriptstyle{\LM}}\right).
\end{align}
Applying Eq.~\eqref{eq:2.9} requires the genuine flow in the low-multiplicity reference, $\lr{v_n^2}_{\scriptscriptstyle{\LM}}$. The LM sample itself cannot supply this, since its flow was assumed to vanish in the construction of the template method. The improved template method instead borrows the flow from a nearby \emph{anchor bin}, denoted ``nextLM'', whose choice depends on the LM choice already made:
\begin{itemize}
\item[(a)] \emph{$pp$ subtraction} (LM $=$ minimum-bias $pp$): the nextLM is the lowest available multiplicity bin of the target system.
\item[(b)] \emph{Peripheral subtraction} (LM $=$ peripheral bin of the target): the nextLM is the next-lowest multiplicity bin.
\end{itemize}
In both cases, the genuine flow in the nextLM bin, $\lr{v_n^2}_{\mathrm{nextLM}}$, is approximated by the template-method value $\lr{v_n^2}^{\mathrm{tmp}}_{\mathrm{nextLM}}$, and Eq.~\eqref{eq:2.9} becomes a one-step correction relating two template-method outputs:
\small{\begin{align}\label{eq:2.10}
  \lr{v_n^2}\! \approx \!\lr{v_n^2}^{\!\mathrm{tmp}}\!-\! f_{2,\mathrm{nextLM}}
  \!\left(\lr{v_n^2}^{\mathrm{tmp}}\! -\!\lr{v_n^2}^{\mathrm{tmp}}_{\scriptscriptstyle{\mathrm{nextLM}}}\right)\;,
\end{align}}\normalsize

where $f_{2,\mathrm{nextLM}} = \lr{v_1^2}^{\obs}/\lr{v_1^2}^{\obs}_{\mathrm{nextLM}}$. ATLAS measurements in $p$+Pb and Pb+Pb collisions show that this correction is modest for $v_2$ and $v_4$ but is important for $v_3$~\cite{ATLAS:2018ngv}.
\end{enumerate}

\vspace*{-0.5cm}
\subsection{Domain of applicability}\label{sec:2.5}

Nonflow subtraction is most reliable at low $\pT$ ($\lesssim 2$~GeV/$c$), where jet-induced correlations on the away side are broad in $\Delta\phi$ and well approximated by a dipolar modulation $\lr{v_1^2}^{\obs} \equiv \lr{\cos\Delta\phi}<0$, which is negative because away-side pair production is biased toward $\Delta\phi \approx \pi$ by momentum conservation. In this regime, the subtraction rests on three conditions of distinct character, stated here for $n\geq 2$:
\begin{align}\label{eq:2.11}
|\lr{v_1^2}^{\nf}|&\approx |\lr{v_1^2}^{\obs}|,\\\label{eq:2.12}
\lr{v_n^2}^{\!\nf}&=a_n|\!\lr{v_1^2}^{\!\nf}\!|,\;\; |\Delta a_n| \ll 1\; ,\\\label{eq:2.13}
\lr{v_n^2}^{\nf}&\ll\lr{v_n^2}^{\obs}.
\vspace*{-0.3cm}
\end{align}

The first condition, Eq.~\eqref{eq:2.11}, makes $\lr{v_1^2}^{\obs}$ a nonflow-dominated estimator (with negative values), since the genuine $v_1$ contribution nearly cancels after $\pT$ integration. The second, Eq.~\eqref{eq:2.12}, requires the nonflow shape to be stable across multiplicity (the change of $a_n$ across $N$ is small $ |\Delta a_n| \ll 1$). This is guaranteed when higher harmonics are negligible ($a_n\ll 1$), as is typical at low $\pT$, but the condition $|\Delta a_n|\ll 1$ is weaker and can hold even when $a_n$ itself is not small. The third condition, Eq.~\eqref{eq:2.13}, by contrast, is a statement about the data: the nonflow contamination of $\lr{v_n^2}^{\obs}$ is small in absolute terms, typically below $20\%$ in $pp$~\cite{ATLAS:2015hzw} and $\pPb$~\cite{ATLAS:2014qaj,Lim:2019cys} at the LHC. When it fails, the subtraction may still be technically valid, but the residual signal becomes too sensitive to the choice of subtraction method to interpret robustly. Together, these conditions keep the nonflow subtraction uncertainty under control.

The situation is very different for correlations involving single hadrons, jets, or heavy-flavor probes at high $\pT$, where the jet-induced correlation in $\LM$, $Y(\Delta\phi)^{\mathrm{LM}}$, is quite narrow in $\Delta\phi$~\cite{ATLAS:2019xqc,ATLAS:2019vcm}. This narrow structure implies sizable harmonics at all orders and a strong $N$-dependence of $a_n$, violating both $|\Delta a_n|\ll 1$ [Eq.~\eqref{eq:2.12}] and Eq.~\eqref{eq:2.13}. The nonflow at $n=2$ often exceeds the genuine signal, so $\lr{v_2^2}$ is small compared to $\lr{v_2^2}^{\nf}$ rather than the other way around. The nonflow systematic uncertainties must therefore be treated with extreme care: if, for example, $\lr{v_2^2}^{\obs} = 10\,\lr{v_2^2}$, a 10\% uncertainty on $\lr{v_2^2}^{\obs}$ produces a 100\% uncertainty on the final extracted $\lr{v_2^2}$. A quantitative assessment via alternative subtraction schemes and/or dedicated Monte Carlo closure tests is needed to evaluate the robustness of the subtraction procedure; such studies have generally not yet been carried out at high $\pT$. A similar dominance $\lr{v_n^2}^{\obs}\gg\lr{v_n^2}$ is also observed for flow in photo-nuclear processes~\cite{ATLAS:2021jhn}.

A qualitatively different failure mode arises in $e^+e^-$ collisions~\cite{Electron-PositronAlliance:2023klx} or in correlations among constituents within a single high-multiplicity jet~\cite{CMS:2023iam}. In both cases the independent-source picture does not apply: all particles originate from one hard interaction rather than a superposition of independent scatterings. In $e^+e^-$, every event is built around a dijet or three-jet topology with $N_s = 2$ or $3$ independent sources, so varying the multiplicity selects different fragmentation configurations of the same hard process, not events with more independent nonflow sources. Inside a single jet, all constituents come from one parton shower ($N_s = 1$), and increasing the jet constituent multiplicity changes the shower properties rather than the source count. In both cases the $1/N^{m-1}$ scaling underlying the LM subtraction breaks down even at low $\pT$, and the method provides no reliable subtraction baseline for collectivity searches in these systems.

\section{Motivation for multi-particle nonflow subtraction}
\label{sec:motivation}

\subsection{Multi-particle observables}
In heavy-ion collisions, event-by-event fluctuations of the collective flow are described by a joint probability distribution $p([\pT], V_1, V_2, \ldots)$, where $[\pT]$ is the event-averaged transverse momentum (which captures the radial flow) and $V_n = v_n e^{i n \Psi_n}$ is the anisotropic flow vector. Experimentally accessible observables are moments of this distribution allowed by global rotational symmetry~\cite{Bhalerao:2011yg,Jia:2014jca},
\begin{align}
\lr{(\delta \pT)^{l} V_{n_1}V_{n_2}\cdots}\,,\quad n_1+n_2+\cdots=0,
\label{eq:3.1}
\end{align}
where $\delta \pT = [\pT]-\lr{[\pT]}$, $l$ is a non-negative integer, and the $n_i$ are non-zero integers. The cumulants measured experimentally are constructed from combinations of these moments.

Examples of such cumulants up to the fourth-order include the following:
\begin{itemize}
\item Two-particle cumulants:\\
$\lr{v_n^2} = \lr{V_{n}V_{-n}}\equiv  \lr{V_{n}V_{n}^*}$ and $\lr{(\delta \pT)^2}$.
\item Three-particle cumulants:\\
$\lr{(\delta\pT)^3}$, $\lr{v_n^2\delta\pT}$, \\
$\lr{\!V_2^2V_4^*\!}$\!, $\lr{\!V_2V_3V_5^*\!}$\!, $\lr{\!V_3^2V_6^*\!}$: asymmetric cumulants.
\item Four-particle cumulants:\\
$c_n\{4\} = \lr{v_n^4} - 2\lr{v_n^2}^2$, $\lr{(\delta\pT)^4} - 3\lr{(\delta\pT)^2}^2$,\\
$\lr{(\delta\pT)^2 v_n^2} - \lr{(\delta\pT)^2}\lr{v_n^2}$,\\
$\mathrm{sc}\!(n,\!m\!)\!=\!\lr{v_n^2 v_m^2}-\lr{v_n^2}\!\lr{v_m^2}$ symmetric cumulants. \\
$\lr{V_2^3V_6^*}$, $\lr{V_3^2V_2^*V_4^*}$... asymmetric cumulants.
\end{itemize}

These cumulants probe the interplay between initial-state geometry fluctuations and the nonlinear hydrodynamic response, and have been central to extracting QGP transport properties from heavy-ion data~\cite{Parkkila:2021tqq}.
\subsection{Limitations of the subevent approach}
The subevent cumulant method~\cite{Jia:2017hbm} suppresses nonflow by selecting particles from pseudorapidity-separated subevents, forcing any residual nonflow contribution to span a large rapidity gap. Take the four-subevent implementation of $c_n\{4\}$ as an example: a dijet can contribute only if its near-side and away-side fragments, together with two additional same-jet particles, populate all four $\eta$ windows. Such configurations become rare as the rapidity gap increases, and the method therefore substantially reduces dijet-related nonflow~\cite{ATLAS:2017rtr,ATLAS:2018ngv,CMS:2019lin,ALICE:2019zfl}.

Suppression, however, is not subtraction. Without an explicit subtraction, the residual nonflow component cannot be quantified. In fact, several published MPC measurements already indicate significant residual nonflow in low-multiplicity $pp$ and $p$+Pb collisions at the LHC:
\begin{itemize}
\item Results for $\lr{v_2^2\delta\pT}$ and $\lr{v_3^2\delta\pT}$ display a non-monotonic multiplicity dependence; moreover, the magnitude and even the sign of these observables vary with the imposed $\eta$ gap~\cite{CMS:2024rvk}.
\item Symmetric cumulants $\mathrm{sc}(2,3)$ and $\mathrm{sc}(2,4)$ exhibit a pronounced dependence on the chosen subevent configuration~\cite{CMS:2017kcs,ATLAS:2018ngv,ALICE:2019zfl,Bilandzic:2013kga} and event-class definition, which leads to inconsistencies among the ATLAS/CMS and ALICE results.
\item Cumulants $c_n\{4\}$ computed in \PYTHIA{}~\cite{Sjostrand:2007gs}, a generator with jet-induced nonflow but no genuine collective flow, change sign and magnitude depending on the $\pT$ selection, event-class definition, and subevent choice. This behavior reflects non-Gaussian fluctuations of nonflow that persist even after subevent suppression~\cite{Jia:2017hbm,ATLAS:2017rtr}.
\end{itemize}

Without an explicit subtraction procedure, these features cannot be uniquely assigned to either flow or nonflow. 

\subsection{Benefits of explicit subtraction}

An explicit nonflow estimate and subtraction in MPC enables: (i) quantification of the residual nonflow systematic in a given measurement; (ii) separation of residual nonflow from flow decorrelation in $\pT$ and $\eta$~\cite{Bozek:2015bna,Jia:2014vja,CMS:2015xmx,ATLAS:2017rij}; (iii) extension of measurements to low multiplicity, where nonflow dominates even after subevent suppression~\cite{ATLAS:2017rtr,ATLAS:2018ngv,ATLAS:2022dov,CMS:2024rvk} and where flow signals are most sensitive to initial-state physics~\cite{Welsh:2016siu}; and (iv) comparison of results from the standard method, by which we mean a cumulant computed over the full acceptance with no subevent separation, and from subevent methods, after nonflow subtraction, as a data-driven consistency test.
% a data-driven consistency check between the standard method and the subevent methods, which should give mutually consistent results after proper subtraction.

There is also an important practical motivation. The four-subevent method reduces the number of unique quadruplets by a factor of $4^4/4! \approx 11$ relative to the standard method~\cite{Jia:2017hbm}; if nonflow can be explicitly subtracted within the standard or two-subevent approach, a substantial statistical gain can be achieved. This improvement is particularly valuable for measurements in high-multiplicity events in small collision systems, where statistical precision is the dominant limiting factor.

For the multi-particle extension developed here we restrict ourselves to low $\pT$, where the correlation structure is GMC-dominated and the genuine flow in the $\LM$ reference can be neglected when the target and reference multiplicities are well separated ($N \gg N_\LM$). In this limit, the template methods reduce to the $c_1$ method of two-particle correlations. We therefore focus on multi-particle extensions of the $c_0$ and $c_1$ scalings, which bracket the systematic range of reasonable nonflow estimates.

\subsection{Domain of applicability}\label{sec:3.4}

The reliability conditions of Sec.~\ref{sec:2.5} carry over directly to multi-particle correlations: the methods developed in this paper are well-founded only at low $\pT$ ($\lesssim 2$~GeV/$c$).  At high $\pT$ both near- and away-side jets fragment into narrow clusters of correlated particles, populating all $m$-particle correlators that involve fragments of the same jet or of the near-/away-side dijet. The resulting jet-induced nonflow contains sizable harmonics at all orders, so all the multi-particle observables discussed above pick up a substantial contamination. The required subtraction grows correspondingly large and becomes acutely sensitive to small uncertainties in the assumed nonflow shape. The factorization of the flow signal into a single one-body distribution is also questionable at high $\pT$, so importing the low-$\pT$ logic into that region may give unphysically large signals, a tension hinted at by the recent CMS $v_2\{4\}$ measurement in $p$+Pb collisions~\cite{CMS:2025kzg}. A positive $\lr{v_2^2}$ value remaining at high $\pT$ after subtraction does not, by itself, establish genuine collective flow.

\section{Subtraction framework}
\label{sec:framework}

So far the reference sample $\LM$ could mean either a low-multiplicity bin of the same collision system or minimum-bias $pp$. We adopt minimum-bias $pp$ as the reference for the remainder of the paper, denoting it by the subscript $\pp$. Using $pp$ avoids the peripheral selection bias of same-system low-multiplicity bins (Appendix~\ref{app:peri} compares the two references directly), but introduces a different complication: the $pp$ multiplicity distribution is broad, and different cumulants weight this distribution differently. The effective multiplicity relevant for each observable must therefore be estimated with care, as we do in Sec.~\ref{sec:Nm}.

\subsection{Scaling principle}
\label{sec:framework1}

Here and below, $\Ocal\{m\}$ denotes a generic $m$-particle cumulant observable, such as $\lr{v_n^2}$ ($m=2$), $\lr{v_n^2\,\dpt}$ ($m=3$), or $c_n\{4\}$ ($m=4$). In the independent-source model, the nonflow contribution to $\Ocal\{m\}$ arises from individually correlated particle production sources. For events with $N_s$ effective sources, each producing a finite number of particles so that $N_s \propto N$, the nonflow contribution scales as~\cite{Borghini:2001vi,Bilandzic:2010jr}
\begin{equation}
  \Ocal\{m\}^{\nf} \propto 1/N_s^{m-1}\propto 1/N^{m-1} \propto |\lr{v_1^2}^{\obs}|^{m-1}\,.
  \label{eq:4.1}
\end{equation}
Equation~\eqref{eq:4.1} is the $m$-particle generalization of the 2PC scaling condition Eq.~\eqref{eq:2.12}: the leading independent-source expectation for the $N$-dependence shared by $m$-particle nonflow observables of a given order. Here $\Ocal\{m\}$ denotes a target $m$-particle cumulant observable. The middle proportionality, $\Ocal\{m\}^{\nf} \propto 1/N^{m-1}$, is the $c_0$-scaling baseline (with $f_1 = N_{\mathrm{LM}}/N$); the right-most proportionality, $\Ocal\{m\}^{\nf} \propto |\lr{v_1^2}^{\obs}|^{m-1}$, is the $c_1$-type refinement that absorbs the residual $N$-dependence from multiplicity-selection bias, which reduces to the $c_0$ form when bias is absent. The simplest choice is an estimator of the same particle order as the target, $k=m$. Such an estimator is not always available and may not be optimal: a lower-order estimator may offer better statistical precision or a more strongly nonflow-dominated reference, so we formulate the subtraction for a general $k$ and account explicitly for the different multiplicity weights below. Heuristically, the $1/N^{m-1}$ scaling follows from counting: of the $\sim N^m$ distinct $m$-tuples in an event, only $\sim N$ are drawn from a single correlated source and thus contribute genuine $m$-particle correlations, so the correlated fraction scales as $N/N^m = 1/N^{m-1}$. Even at low $\pT$, deviations from Eqs.~\eqref{eq:4.1} and~\eqref{eq:2.12} can arise from multiplicity-selection biases, initial-state saturation effects~\cite{Dusling:2015gta,Schlichting:2016sqo}, and final-state string interactions such as shoving~\cite{Bierlich:2020naj}.

\subsection{Effective event multiplicity}
\label{sec:Nm}

When minimum-bias $pp$ collisions are used as the nonflow reference, a technical subtlety arises: the $pp$ multiplicity distribution is broad (reaching $> 100$ at LHC energies). Because an $m$-particle correlator is weighted event by event by the number of available $m$-particle combinations, high-$N$ events contribute far more strongly than low-$N$ events. The effective multiplicity must therefore use the same combination weights rather than simple event weights. We denote this combination-weighted multiplicity by $N_m$, which differs from the event average $\lr{N}$.

We define $N_m$ from the combination-weighted average of $1/N$:
\begin{align}
  \frac{1}{N_m}
  &\equiv \lr{\frac{1}{N}}_{\!\mathrm{comb}}
  \nonumber\\
  &= \frac{\sum_i 1 /N_i \times N_i(N_i-1)\cdots(N_i-m+1)}
         {\sum_i N_i(N_i-1)\cdots(N_i-m+1)},
  \label{eq:4.2}
\end{align}
which for $N_i \gg m$ (see Appendix~\ref{app:Nn}) simplifies to
\begin{equation}
  N_m
  \approx \frac{\sum_i N_i^m}{\sum_i N_i^{m-1}}.
  \label{eq:4.3}
\end{equation}
In this definition, the order $m$ enters through the combination weight. Thus, $N_m$ is the inverse of the combination-weighted mean of $1/N$, when $N\gg m$. Note that $N_1\equiv\lr{N}$ and $N_2\approx\lr{N^2}/\lr{N}$. For the multiplicity distributions considered here, $N_m$ increases with $m$ and satisfies $N_m \geq \lr{N}$.

The alternative is to average $1/N^{m-1}$ instead of $1/N$. This gives an effective multiplicity $M_m$ that is exact if the $1/N^{m-1}$ scaling holds event by event within the sample, and it may therefore seem more natural (Appendix~\ref{app:Mn}). The two definitions coincide for $m=2$, while $M_m\leq N_m$ for $m>2$. In the narrow target bins used here, $M_m\simeq N_m\simeq\lr{N}$. The reference factor also cancels when the target and estimator have the same order, but it matters for pure multiplicity scaling and mixed-order estimators. We use $N_m$ for two reasons. Its average of $1/N$ is less sensitive than the average of $1/N^{m-1}$ to the low-$N$ part of the broad $pp$ distribution, where the assumed scaling is least reliable (Sec.~\ref{subsec:scaling}). Tests with \HIJING{} also show better closure in many of the mixed-order cases studied (Appendix~\ref{app:Mn}). The difference between the two choices is part of the model dependence and should be tested in data before settling on one versus the other.

The $1/N^{m-1}$ scaling of $\Ocal\{m\}$ can be verified by plotting $|\Ocal\{m\}|^{1/(m-1)}$ against $N_m$~\footnote{We use absolute value, since the measured estimators $\lr{v_1^2}^{\obs}$ and $\lr{v_1^2\,\dpt}^{\obs}$ are negative.}. What is plotted on the $y$-axis is the measured $\Ocal\{m\}^{\obs}$ raised to that power, not the approximate form of Eq.~\eqref{eq:4.1}. This implies that the well-known centrality-bin-width correction (CBWC)~\cite{Luo:2013bmi} must be applied \emph{both} to the observable ($y$-axis) \emph{and} to the effective multiplicity ($x$-axis). For studies that only address the qualitative centrality dependence of a single observable, the distinction between $N_m$ and $\lr{N}$ is a mild change in the $x$-values, and has traditionally been neglected. The distinction matters, however, when quantitatively checking the scaling behavior or performing the peripheral subtraction, where the $x$-value enters the calculation directly~\footnote{We note that the CBWC correction for $\lr{N}$ should also be applied in 2PC analyses, but rarely is.}.

Note that the exact formula Eq.~\eqref{eq:4.2} should be used if $N$ is small. For the subevent method, the weight is modified. For example, the two-subevent three-particle correlator has weight $N_a N_b(N_a + N_b - 2)$ rather than $N(N-1)(N-2)$; $N_m$ must be recalculated accordingly. This recalculation was performed separately for every standard-method and two-subevent result in this work; the resulting reference values are listed in Table~\ref{tab:2}, where the two methods differ by less than $2\%$ at $\snn=5.36$~TeV. The conceptual definition and consequences in Eqs.~(\ref{eq:4.2})--(\ref{eq:4.3}) remain unchanged.

In data, $N_m$ should be constructed from the efficiency-corrected charged-particle multiplicity distribution using the same event, track, and subevent selections as the correlator. For a constant, independent single-particle efficiency $\epsilon$, and sufficiently large multiplicities at fixed correlator order, $N_m^{\mathrm{obs}}\approx\epsilon N_m^{\mathrm{true}}$. The common efficiency therefore cancels to leading order in $N_{m,\pp}/N_{k,\pp}$, but finite-multiplicity corrections can remain even for independent track detection. These corrections, together with two-track resolution, trigger effects, and event-selection correlations, should be assessed and propagated through $W_{m,k}$.

\subsection{Nonflow scaling estimators: correlators containing $v_1$}
\label{sec:v1refs}

A useful scaling estimator must itself be dominated by nonflow. The natural choices are cumulants containing $v_1$, for two reasons:
\begin{itemize}
\item At low $\pT$, a rapidity gap suppresses short-range nonflow from near-side jets, resonance decays, and HBT. The surviving
long-range nonflow includes away-side jet correlations and the dipolar correlation induced by global momentum conservation (GMC), motivating estimators containing $v_1$~\cite{Borghini:2003ur}. The leading GMC contribution need not survive in every harmonic
projection. For example, for an otherwise azimuthally isotropic single-particle distribution, the $1/N^2$ GMC contribution to
$C_{123}$ vanishes. Away-side dijet correlations can still contribute to $C_{123}$, with an expected $1/N^2$ scaling in the independent-source picture.
  
\item The genuine dipolar flow $v_1$ changes sign at $\pT \approx \lr{\pT} \sim 0.5$--$1.0$~GeV/$c$~\cite{Borghini:2002mv,Luzum:2010fb,Teaney:2010vd,ATLAS:2012at}. When integrated over the full $\pT$ range, the genuine $v_1$ contribution to the correlator is expected to be small but may not vanish. This residual contamination should be assessed by varying the $\pT$ interval and comparing alternative estimators. We leave this quantitative study to future work.
\end{itemize}

The relevant scaling estimators and their expected independent-source nonflow scalings include, but are not limited to, the following:
\small{
\begin{align}
  |\lr{v_1^2}^{\obs}| &\sim 1/N, \label{eq:4.4}\\
  |\lr{v_1^2\,\dpt}^{\obs}| &\sim 1/N^2, \label{eq:4.5}\\
  c_1\{4\}^{\obs} = \lr{v_1^4}^{\obs} - 2\bigl(\lr{v_1^2}^{\obs}\bigr)^2 &\sim 1/N^3,
  \label{eq:4.6}\\
  C_{112} \equiv \lr{V_1 V_1 V_2^*}^{\obs} &\sim 1/N^2, \label{eq:4.7}\\
  C_{123} \equiv \lr{V_1 V_2 V_3^*}^{\obs} &\sim 1/N^2, \label{eq:4.8}\\
  {\mathrm{sc}}(1,n)^{\obs} \equiv \lr{v_1^2 v_n^2}^{\obs} - \lr{v_1^2}^{\obs}\lr{v_n^2}^{\obs} &\sim 1/N^3. \label{eq:4.9}
\end{align}}\normalsize

\subsection{General subtraction formula}
\label{sec:general_sub}
%This assumption is not tested by the present nonflow-only \HIJING{} study and requires validation in samples containing both flow and nonflow.
We assume that the observed cumulant can be written as the sum of a genuine flow contribution and an additive nonflow contribution. Mixed flow--nonflow terms are assumed to be negligible or to follow the same reference scaling as the nonflow contribution.

The general nonflow subtraction formula is
\begin{widetext}
\begin{equation}
  \Ocal\{m\} = \Ocal\{m\}^{\obs} - f_{\Ocal}^{\Ecal} \times \Ocal\{m\}^{\obs}_\pp  = \Ocal\{m\}^{\obs}
    - \left(\left(\frac{\Ecal\{k\}^{\obs}}{\Ecal\{k\}^{\obs}_\pp}\right)^{1/(k-1)}
     \underbrace{\frac{N_{m,\pp}}{N_{k,\pp}} \cdot \frac{N_k}{N_m}}_{W_{m,k}}\right)^{m-1}
    \Ocal\{m\}^{\obs}_\pp ,
  \label{eq:4.10}
\end{equation}
\end{widetext}
where $\Ocal\{m\}$ is the target $m$-particle cumulant, $\Ecal\{k\}$ is a $k$-particle cumulant dominated by nonflow that serves as the scaling estimator, and the subscript $\pp$ denotes the minimum-bias $pp$ baseline. The \emph{multiplicity-reweighting factor}
\begin{equation}
  W_{m,k} \equiv \frac{N_{m,\pp}}{N_{k,\pp}} \cdot \frac{N_k}{N_m}\approx \frac{N_{m,\pp}}{N_{k,\pp}}
  \label{eq:4.11}
\end{equation}
accounts for the fact that target and reference observables are weighted by different numbers of multiplets, and satisfies $W_{m,m}=1$ and $W_{m,k} = 1/W_{k,m}$. $W_{m,k} \approx 1$ for narrow multiplicity bins but deviates significantly for broad multiplicity distributions. Since events for the system of interest are always binned in narrow multiplicity ranges, $N_k\approx N_m\approx \lr{N}$, but this is not the case when the entire minimum-bias $pp$ ensemble is used as one bin. In Eq.~\eqref{eq:4.10}, the estimator ratio enters with power $(m-1)/(k-1)$, whereas $W_{m,k}$ enters with power $m-1$. The latter amplifies modest deviations of $W$ from unity into large subtraction errors for higher-order target cumulants. Omitting $W_{m,k}$ is a frequent failure mode of multi-particle nonflow subtraction, as demonstrated in Sec.~\ref{sec:res3}.

%With $N_m$, the $N$-dependence of these ratios is unchanged but their normalization shifts by the constant factor $M_{m,\pp}/N_{m,\pp}$ for $m> 2$, which in \HIJING{} this choice brings the ratios closer to unity (Appendix~\ref{app:Mn}).
The residual fractional nonflow after subtraction is
\begin{equation}
  \frac{\Ocal\{m\}}{\Ocal\{m\}^{\obs}}
  = 1 - \left(\frac{R_\Ecal}{R_\Ocal}\right)^{m-1},
  \label{eq:4.12}
\end{equation}
where we introduce the dimensionless ratio
\small{\begin{align}
  R_\Ecal \!\equiv\! \frac{N_k}{N_{k,\pp}}\!\left(\!\frac{\Ecal\{k\}^{\obs}}{\Ecal\{k\}^{\obs}_\pp}\!\right)^{\!1/(k-1)},  R_\Ocal\!\equiv \! \frac{N_m}{N_{m,\pp}}\!\left(\!\frac{\Ocal\{m\}^{\obs}}{\Ocal\{m\}^{\obs}_\pp}\!\right)^{\!1/(m-1)}.\label{eq:4.13}
\end{align}}\normalsize
If nonflow scales exactly as $1/N^{m-1}$, then $R_\Ocal$ and $R_\Ecal$ are close to unity and independent of target multiplicity. The ratio $D \equiv R_\Ocal/R_\Ecal$ measures how closely the reference observable tracks the target nonflow: perfect subtraction has $D = 1$.

Under the same additive assumption, we propose the following template extensions to account for genuine flow in the $pp$ reference sample by modifying Eq.~\eqref{eq:4.10}:
\begin{align}\label{eq:4.14}
&\Ocal\{m\}^{\mathrm{tmp}} = \frac{\Ocal\{m\}^{\obs} - f_{\Ocal}^{\Ecal} \,\Ocal\{m\}^{\obs}_\pp}{1-f_{\Ocal}^{\Ecal}}\\\nonumber
&\Ocal\{m\} = \Ocal\{m\}^{\mathrm{tmp}} \\\label{eq:4.15}
&\hspace*{0.5cm}- f_{\Ocal,\mathrm{nextLM}}^{\Ecal}(\Ocal\{m\}^{\mathrm{tmp}}-\Ocal\{m\}^{\mathrm{tmp}}_{\mathrm{nextLM}})
\end{align}
where ``nextLM'' denotes the lowest multiplicity bin in the system of interest. These equations are defined in direct analogy to Eqs.~\eqref{eq:2.7} and~\eqref{eq:2.10}. These template-method extensions are not exercised in the present analysis; their validation in samples containing both genuine flow and nonflow is left for future work.

\subsection{Classification of methods}
\label{sec:methods}
As argued in Sec.~\ref{sec:motivation}, we focus on the multi-particle extension of the $c_0$ and $c_1$ scalings, which bracket the systematic range of reasonable nonflow estimates. We identify 
one $c_0$-scaling and three $c_1$-based variants according to the choice of $\Ecal\{k\}$.

\subsubsection{Method 1: $c_0$ scaling}

When $\Ecal\{k\} = 1/N^{k-1}$ (pure multiplicity scaling) is used, Eq.~\eqref{eq:4.10} takes the form
\begin{align}\nonumber
\lr{v_n^2} &= \lr{v_n^2}^{\obs}- N_{2,\pp}/N_2 \,\lr{v_n^2}_\pp^{\obs},\\\nonumber
\lr{v_n^2\,\dpt} &= \lr{v_n^2\,\dpt}^{\obs}-(N_{3,\pp}/N_3)^2 \,\lr{v_n^2\,\dpt}_\pp^{\obs},\\
c_n\{4\}&= c_n\{4\}^{\obs}-(N_{4,\pp}/N_4)^3 \,c_n\{4\}_\pp^{\obs}.\label{eq:4.16}
\end{align}

\subsubsection{Method 2a: native estimator}

Use a $v_1$-containing reference observable in Eq.~\eqref{eq:4.10} matched at the same order as the target, e.g. $m=k$:
\begin{align}\nonumber
  \lr{v_n^2}  &= \lr{v_n^2}^{\obs}- \frac{\lr{v_1^2}^{\obs}}{\lr{v_1^2}^{\obs}_\pp}\lr{v_n^2}_\pp^{\obs},\\\nonumber
  \lr{v_n^2\,\dpt}  &= \lr{v_n^2\,\dpt}^{\obs} -\frac{\lr{v_1^2\,\dpt}^{\obs}}{\lr{v_1^2\,\dpt}^{\obs}_\pp}\lr{v_n^2\,\dpt}_\pp^{\obs},\\\nonumber
  c_n\{4\}  &= c_n\{4\}^{\obs}- \frac{c_1\{4\}^{\obs}}{c_1\{4\}^{\obs}_\pp}c_n\{4\}^{\obs}_\pp.\\\label{eq:4.17}
 \mathrm{sc}(n,m)  &= \mathrm{sc}(n,m)^{\obs} - \frac{\mathrm{sc}(1,n)^{\obs}}{\mathrm{sc}(1,n)^{\obs}_\pp} \mathrm{sc}(n,m)^{\obs}_\pp. 
\end{align}

\subsubsection{Method 2b: pseudo-native estimator}

For mixed-harmonic targets such as $\lr{V_2^2 V_4^*}$ or $\lr{V_2 V_3 V_5^*}$, one may use a pseudo-native estimator with $m=k$ but containing harmonics of different order. For example
\begin{equation}
  \lr{V_2^2 V_4^*}
  = \lr{V_2^2 V_4^*}^{\obs}
    - \frac{\lr{V_1^2 V_2^*}^{\obs}}
           {\lr{V_1^2 V_2^*}^{\obs}_\pp}
      \lr{V_2^2 V_4^*}^{\obs}_\pp,
  \label{eq:4.18}
\end{equation}
or alternatively using $\lr{V_1 V_2 V_3^*}^{\obs}$ as the estimator. The harmonic indices of the estimator must sum to zero [Eq.~\eqref{eq:3.1}]. The seemingly analogous $\lr{V_1^2V_4^*}$ is therefore not a valid observable: $1+1-4\neq 0$, so it averages to zero over the global azimuthal orientation.

\subsubsection{Method 2c: mixed-order estimator}

Estimate the nonflow using a reference of different order, e.g. $k\neq m$; for example:
\small{\begin{align}\nonumber
  \lr{v_n^2} & = \lr{v_n^2}^{\obs}- \left(\frac{\lr{v_1^2\,\dpt}^{\obs}}{\lr{v_1^2\,\dpt}^{\obs}_\pp} \right)^{1/2} \frac{N_{2,\pp}}{N_{3,\pp}}\frac{N_3}{N_2} \lr{v_n^2}_\pp^{\obs},\\\nonumber
  \lr{v_n^2\,\dpt} & = \lr{v_n^2\,\dpt}^{\obs}- \left(\frac{\lr{v_1^2}^{\obs}}{\lr{v_1^2}^{\obs}_\pp} \frac{N_{3,\pp}}{N_{2,\pp}}\frac{N_2}{N_3} \right)^2 \lr{v_n^2\,\dpt}_\pp^{\obs},\\\nonumber
  c_n\{4\} &= c_n\{4\}^{\obs}- \left(\frac{\lr{v_1^2}^{\obs}}{\lr{v_1^2}^{\obs}_\pp} \frac{N_{4,\pp}}{N_{2,\pp}}\frac{N_2}{N_4} \right)^3 c_n\{4\}_\pp^{\obs},\\\label{eq:4.19}
  c_n\{4\} &= c_n\{4\}^{\obs}-\!\left(\!\frac{\lr{v_1^2\,\dpt}^{\obs}}{\lr{v_1^2\,\dpt}^{\obs}_\pp}\!\right)^{3/2}\hspace*{-0.2cm}\left(\frac{N_{4,\pp}}{N_{3,\pp}}\frac{N_3}{N_4} \right)^{3}\hspace*{-0.2cm} c_n\{4\}_\pp^{\obs},
\end{align}}\normalsize
The problem is that there are many possible target-reference combinations. It is not clear a priori which ones are good, so they need to be carefully validated in model studies.

\subsubsection{Considerations in subtraction strategy}

The nonflow scaling argument in Sec.~\ref{sec:framework1} applies to multi-particle cumulants rather than to multi-particle moments; the two have fundamentally different scaling behavior. For example, the four-particle moment $\lr{v_n^4}$ contains disconnected products of two pairwise nonflow correlations; each pair scales as $1/N$, so their product scales as $1/N^2$. The cumulant $c_n\{4\}$ removes these disconnected products, leaving the connected contribution that scales as $1/N^3$ in the independent-source picture. Nonflow subtraction must therefore be applied at the cumulant level throughout.

Second, under the independent-source scaling assumption, the multiplicity-reweighting factor is required whenever the reference observable and the target observable do not have the same order. Omitting $N_{m,\pp}/N_{k,\pp}$ in Eq.~\eqref{eq:4.10} leads to large, order-dependent errors (Sec.~\ref{sec:res3}). If the subtraction still over- or undersubtracts once the factor is included, that means the independent-source assumption is imperfect somewhere, not that the reweighting was unnecessary.

Third, given the large number of possible target--reference combinations, it is important to check their performance explicitly. The best choice of $\Ecal\{k\}$ depends on the dynamical origin of the target nonflow and must be determined case by case in model studies and real data analyses.

For this first study we restrict ourselves to three representative observables: $\lr{v_2^2}$, $\lr{v_2^2\,\dpt}$, and $c_2\{4\}$. Together they cover the central physics goals of the small-system flow program and exercise all the technical aspects of the subtraction. The three serve as a testbed for evaluating subtraction performance, comparing alternative strategies, and picking out the best-performing reference for each target.

\section{Analysis setup}
\label{sec:analysis}

We use the \HIJING{}~1.383 generator~\cite{Gyulassy:1994ew}, which includes minijet production via pQCD-inspired multi-parton scattering and string fragmentation, but no genuine collective flow. Five datasets are generated, each with $2\times 10^9$ minimum-bias events:
\begin{itemize}
\item $pp$ and $\OO$ at $\snn = 5.36$~TeV,
\item $pp$, $\OO$, and $\dAu$ at $\snn = 200$~GeV.
\end{itemize}

Charged particles are selected with $0.2 < \pT < 2.0$~GeV/$c$ and $|\eta| < 2.5$ at LHC energy (ATLAS/CMS acceptance) or $|\eta| < 1.5$ at RHIC energy (STAR acceptance). The total charged and neutral particle multiplicity $N$ is counted within these acceptances.

Observables are computed with both the standard method (over the full $\eta$ acceptance defined above) and the two-subevent method, in which the two subevents are defined by $\eta_a < -0.5$ and $\eta_b > 0.5$, giving a minimum gap $|\Delta\eta|_{\min} = 1.0$.

The three target observables $\lr{v_2^2}$, $\lr{v_2^2\,\dpt}$, and $c_2\{4\}$ are computed using the generic cumulant framework~\cite{Bilandzic:2013kga}. They are evaluated in unit-multiplicity bins~\cite{Luo:2013bmi} and subsequently rebinned into wider multiplicity ranges. 

For each target, three scaling estimators, $1/N^{k-1}$, $\lr{v_1^2}^{\obs}$, and $\lr{v_1^2\,\dpt}^{\obs}$, are used to track the $N$-dependence of the nonflow. This gives nine target--estimator combinations in total, summarized in Table~\ref{tab:1}. The explicit expressions for these combinations are given as part of Eqs.~\eqref{eq:4.16}--\eqref{eq:4.19}.

The $1/N^{m-1}$ baseline follows from the independent-source picture rather than from \HIJING{}. However, the departure from that baseline, including the estimator ranking and $K_{\Ocal}^{\Ecal}(N)$ [Eq.~\eqref{eq:6.2}], is model dependent, and depends on the composition and rapidity structure of nonflow~\cite{Schenke:2022mjv}. A second generator would test the transferability of those quantities.

\begin{figure*}[tp]
\centering
\includegraphics[width=1\linewidth]{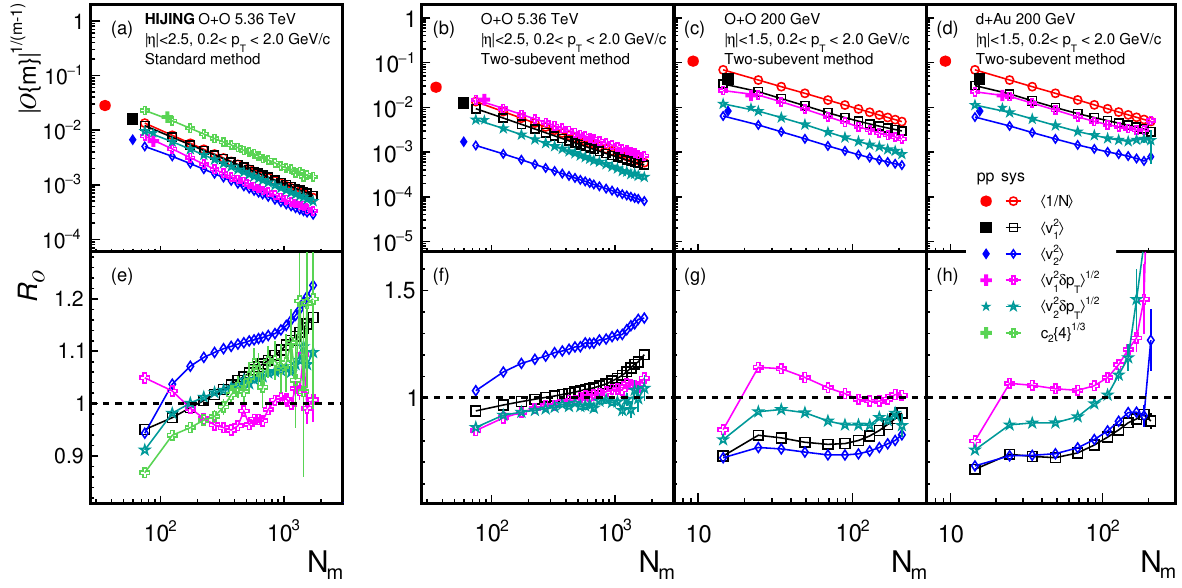}
\caption{\textbf{Multiplicity scaling behavior of observables}. Observables $|\Ocal\{m\}|^{1/(m-1)}$ (top row) and their multiplicity-scaled version $R_\Ocal$ (bottom row) as a function of effective multiplicity $N_m$ in the \HIJING{} model. Results are obtained using the standard method in O+O at $\snn = 5.36$~TeV (left column), and the two-subevent method in O+O at $\snn = 5.36$~TeV (2nd column), O+O at $\snn = 200$~GeV (third column), and $d$+Au at $\snn = 200$~GeV (right column).  Solid symbols show the corresponding minimum-bias $pp$ results at the same energy, plotted at $N_{m,\pp}$. Deviations of $R_\Ocal$ from unity reflect both departures from independent-source scaling and the broad-reference averaging prescription. For observables that are always negative ($\lr{v_1^2}^{\obs}$ and $\lr{v_1^2\,\dpt}^{\obs}$), the absolute values are plotted. The $c_2\{4\}$ observable is shown only for the standard method, because in the two-subevent method its values are strongly suppressed and statistics-limited.}
\label{fig:1}
\end{figure*}

\begin{table}[!t]
\centering
\caption{Subtraction scale factors for each target observable under the three scaling estimators of Eqs.~\eqref{eq:4.16}--\eqref{eq:4.19}. The estimator ratios enter with power $(m-1)/(k-1)$ and the multiplicity-reweighting factors $W_{m,k}$ with power $m-1$, following Eq.~\eqref{eq:4.10}.}
\label{tab:1}
\begin{tabular}{lccc}
\hline\hline
                        & $c_0$-scaling     & $c_1$-scaling & $\lr{v_1^2\,\dpt}$-scaling \\\hline\\
$\lr{v_2^2}$           & $\frac{N_{2,\pp}}{N_2}$     & $\frac{\lr{v_1^2}^{\obs}}{\lr{v_1^2}^{\obs}_\pp}$   & $\left(\!\frac{\lr{v_1^2\,\dpt}^{\obs}}{\lr{v_1^2\,\dpt}^{\obs}_\pp}\!\right)^{\!1/2} \hspace*{-0.3cm}W_{2,3}$ \\
&&&\\
$\lr{v_2^2\,\dpt}$     & $\frac{N^2_{3,\pp}}{N^2_3}$ & $\left(\!\frac{\lr{v_1^2}^{\obs}}{\lr{v_1^2}^{\obs}_\pp} W_{3,2}\!\right)^2$ & $\frac{\lr{v_1^2\,\dpt}^{\obs}}{\lr{v_1^2\,\dpt}^{\obs}_\pp}$      \\
&&&\\
$c_2\{4\}$                       & $\frac{N^3_{4,\pp}}{N^3_4}$ & $\left(\!\frac{\lr{v_1^2}^{\obs}}{\lr{v_1^2}^{\obs}_\pp} W_{4,2}\!\right)^3$ & $\left(\!\frac{\lr{v_1^2\,\dpt}^{\obs}}{\lr{v_1^2\,\dpt}^{\obs}_\pp}\!\right)^{\!3/2}\hspace*{-0.3cm} W_{4,3}^3$ \\
\hline\hline
\end{tabular}
\end{table}

\section{Results}\label{sec:res}
\subsection{Nonflow scaling behavior}\label{sec:res1}
Before applying the subtraction, we first verify the $1/N^{m-1}$ scaling for both target and reference observables. 

A note on what \HIJING{} tests. Because \HIJING{} contains no genuine flow, $\lr{v_n^2}=0$ and $\lr{v_n^2}^{\obs}=\lr{v_n^2}^{\nf}$; the data-quality condition Eq.~\eqref{eq:2.13}, $\lr{v_n^2}^{\nf}\ll\lr{v_n^2}^{\obs}$, is trivially violated. \HIJING{} therefore stress-tests the $1/N^{m-1}$ scaling of nonflow [Eq.~\eqref{eq:4.1}], the dipolar-shape assumption [Eq.~\eqref{eq:2.12}], and the multiplicity-reweighting machinery. It does not test recovery of a known flow signal in the presence of nonflow; that requires a flow-afterburner or another generator which we defer to a future study. Residual fractions quoted here are normalized to the observed nonflow-only value in \HIJING{}. The actual bias on measurements depends on how this residual nonflow compares to the magnitude of genuine flow signal. Unless explicitly stated, the target observables plotted and discussed below are the observed ones, and we drop the $\obs$ label on them where no confusion arises (except for the $v_1$-containing obervables).

Figure~\ref{fig:1}(a) displays $|\Ocal\{m\}|^{1/(m-1)}$ from the standard method at the LHC energies. The results from O+O and $pp$ collisions are represented by the open symbols and filled symbols, respectively. The O+O results are calculated differentially as a function of $N_m$, while those in $pp$ are calculated over all events and shown as a single data point at the corresponding $N_{m,\pp}$, obtained from the same combination-weighted prescription of Eq.~\eqref{eq:4.2} and therefore different for each correlator order. These values are listed in Table~\ref{tab:2}.

All six observables, $1/N$, $\lr{v_1^2}^{\obs}$, $\lr{v_1^2\,\dpt}^{\obs}$, $\lr{v_2^2}$, $\lr{v_2^2\,\dpt}$, and $c_2\{4\}$ approximately follow the $1/N^{m-1}$ scaling in O+O and $pp$ collisions. The $pp$ data must be plotted at $N_{m,\pp}$; otherwise they would not align with the O+O scaling for the corresponding observable, and any subtraction based on them would be strongly biased.

Figure~\ref{fig:1}(e) displays the corresponding $R_\Ocal$ ratios from Eq.~\eqref{eq:4.13}. These ratios cluster around unity with deviations of up to $20$--$30\%$ depending on the observable. By construction, $R_{1/N} = 1$ while the deviations for other observables reflect their genuine departure from the $1/N$-scaling. The $\lr{v_1^2}^{\obs}$ ratio closely tracks $\lr{v_2^2}$, supporting its use as the scaling estimator in the $c_1$ method. Much of the nonflow is already absorbed by the simple $1/N$ scaling.

Results from the two-subevent method appear in the right three columns for O+O at $\snn=5.36$~TeV, O+O at $\snn=200$~GeV, and $d$+Au at $\snn=200$~GeV, respectively. The top panels show qualitatively similar scaling behavior, but the $R_\Ocal$ ratios reveal quantitative dependence on collision energy and system. For O+O at $5.36$~TeV, the $R_\Ocal$ behavior agrees within $10\%$ between the standard method and two-subevent methods (1st and 2nd columns). The comparison between $5.36$~TeV and $200$~GeV in O+O (2nd and 3rd columns) reveals a qualitative difference: $R_\Ocal$ for $\Ocal=\lr{v_1^2}^{\obs}$ and $\lr{v_2^2}$ lies above unity at $5.36$~TeV (decreasing slower than $1/N$) and below unity at $200$~GeV (decreasing faster than $1/N$). The scaling is similar in O+O and $d$+Au at $200$~GeV, although the departure from $1/N$ tends to be somewhat larger in central $d$+Au, possibly because of the stronger multiplicity-selection bias arising from the steeper $p(N)$ distribution. 

This variation of $R_\Ocal$ across observables motivates the search for the best nonflow estimator $\Ecal\{k\}$ for each target. The overall behavior of $R_\Ocal$ separates into two groups: $\Ocal \in \{1/N,\,\lr{v_1^2}^{\obs},\,\lr{v_2^2}\}$ on the one hand, and $\Ocal \in \{\lr{v_1^2\,\dpt}^{\obs},\,\lr{v_2^2\,\dpt}\}$ on the other. Pairs within the same group provide a better match between target observable and scaling estimator.

\begin{figure*}[tp]
\centering
\includegraphics[width=1\linewidth]{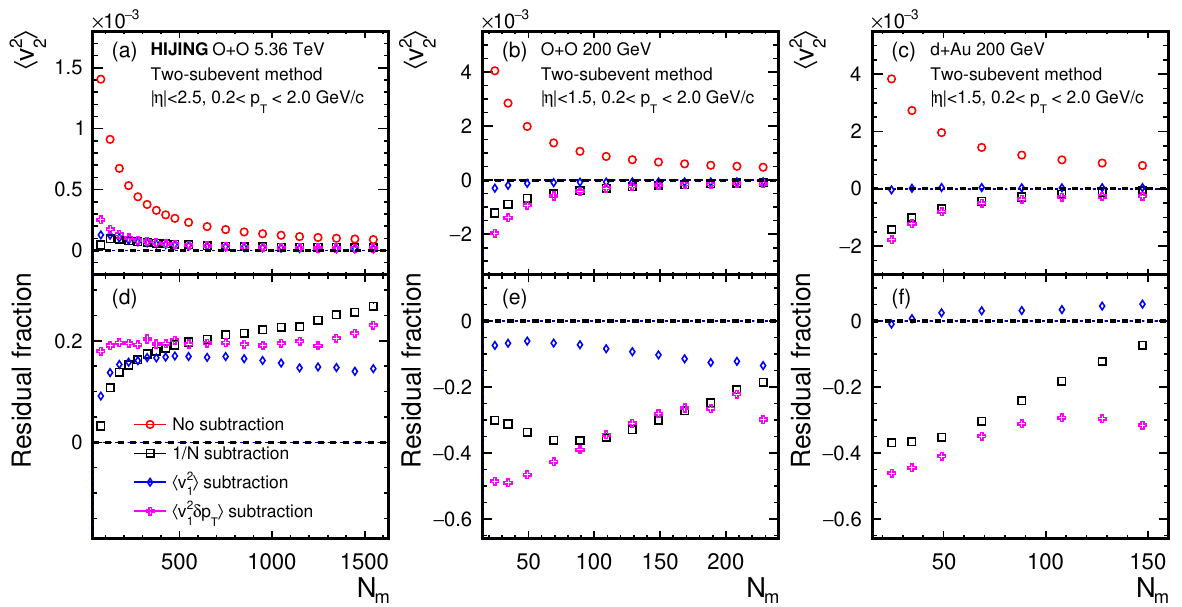}
\caption{\textbf{Nonflow subtraction for $\lr{v_2^2}$}. Performance for O+O collisions at 5.36 TeV (left), O+O collisions at 200 GeV (middle), and $d$+Au collisions at 200 GeV (right) in two-subevent method as a function of $N_m$. The top panels show $\lr{v_2^2}$ before and after subtraction. The bottom panels show the fractional residual nonflow defined in Eq.~\eqref{eq:4.12} using $1/N$-scaling (open boxes), $\lr{v_1^2}^{\obs}$-scaling (open diamonds), and $\lr{v_1^2\,\dpt}^{\obs}$-scaling (open crosses).}
\label{fig:2}
\end{figure*}
\subsection{Subtraction performance}\label{sec:res2}
Because the scaling behavior varies across observables, the performance of nonflow subtraction depends on the target--reference combination. We compare the three reference scalings ($1/N$, $\lr{v_1^2}^{\obs}$, and $\lr{v_1^2\,\dpt}^{\obs}$) for each of the three target observables ($\lr{v_2^2}$, $\lr{v_2^2\,\dpt}$, and $c_2\{4\}$) in three collision systems, a $3\times 3\times 3$ matrix of cases. 

Throughout this section, the residual fraction is the subtracted value divided by the observed \HIJING{} value, $\Ocal\{m\}/\Ocal\{m\}^{\obs}$, as defined in Eq.~\eqref{eq:4.12}. Zero denotes closure, a positive value denotes undersubtraction, a negative value denotes oversubtraction, and a value near unity means that little nonflow has been removed. Figure~\ref{fig:2} shows the performance for $\lr{v_2^2}$ using three scaling ans\"atze in three collision systems. The nonflow contribution, already suppressed by the two-subevent requirement, is further reduced by the explicit subtraction. For O+O at $5.36$~TeV the nonflow is undersubtracted, leaving a positive residual fraction; for O+O and $d$+Au at $200$~GeV it is mostly oversubtracted, leaving a negative one. Across all three systems the $\lr{v_1^2}^{\obs}$-scaling method achieves the best performance, with residual nonflow within $15\%$. The $1/N$ scaling performs worst at LHC energy, while $\lr{v_1^2\,\dpt}^{\obs}$-scaling performs worst at RHIC energy. The present study does not isolate a unique origin of these energy-dependent differences, reflecting contributions from both multiplicity-selection effects and changes in the composition and correlation structure of nonflow.

\begin{figure*}[tp]
\centering
\includegraphics[width=1\linewidth]{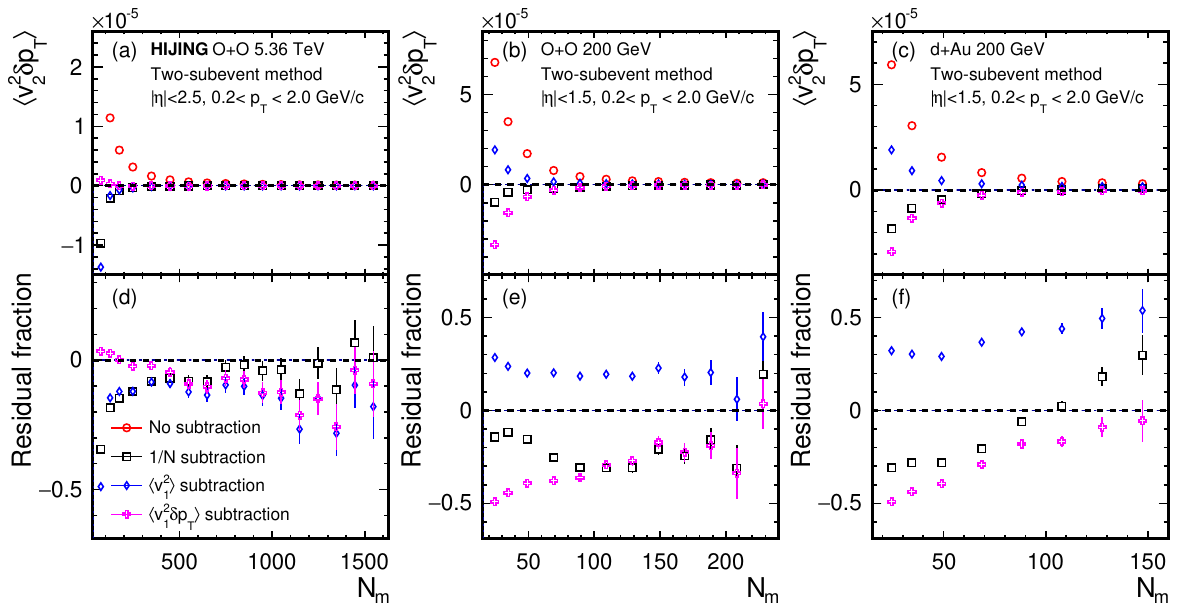}
\caption{\textbf{Nonflow subtraction for $\lr{v_2^2\,\dpt}$}. Performance for O+O collisions at 5.36 TeV (left), O+O collisions at 200 GeV (middle), and $d$+Au collisions at 200 GeV (right) in two-subevent method as a function of $N_m$. The top panels show $\lr{v_2^2\,\dpt}$ before and after subtraction. The bottom panels show the fractional residual nonflow defined in Eq.~\eqref{eq:4.12} using $1/N$-scaling (open boxes), $\lr{v_1^2}^{\obs}$-scaling (open diamonds), and $\lr{v_1^2\,\dpt}^{\obs}$-scaling (open crosses).}
\label{fig:3}
\end{figure*}
Figure~\ref{fig:3} shows the performance for $\lr{v_2^2\,\dpt}$. Across the systems studied, the $1/N$ scaling gives the smallest residuals, below $20\%$ outside the lowest-multiplicity bins. The $\lr{v_1^2}^{\obs}$ and $\lr{v_1^2\,\dpt}^{\obs}$ scalings leave larger, energy-dependent residuals and do not faithfully track the $N$-dependence of the target nonflow. The subtractions nevertheless remain improvements over no subtraction at all in the intermediate- and high-multiplicity regions.

\begin{figure}[htb]
\centering
\includegraphics[width=0.8\linewidth]{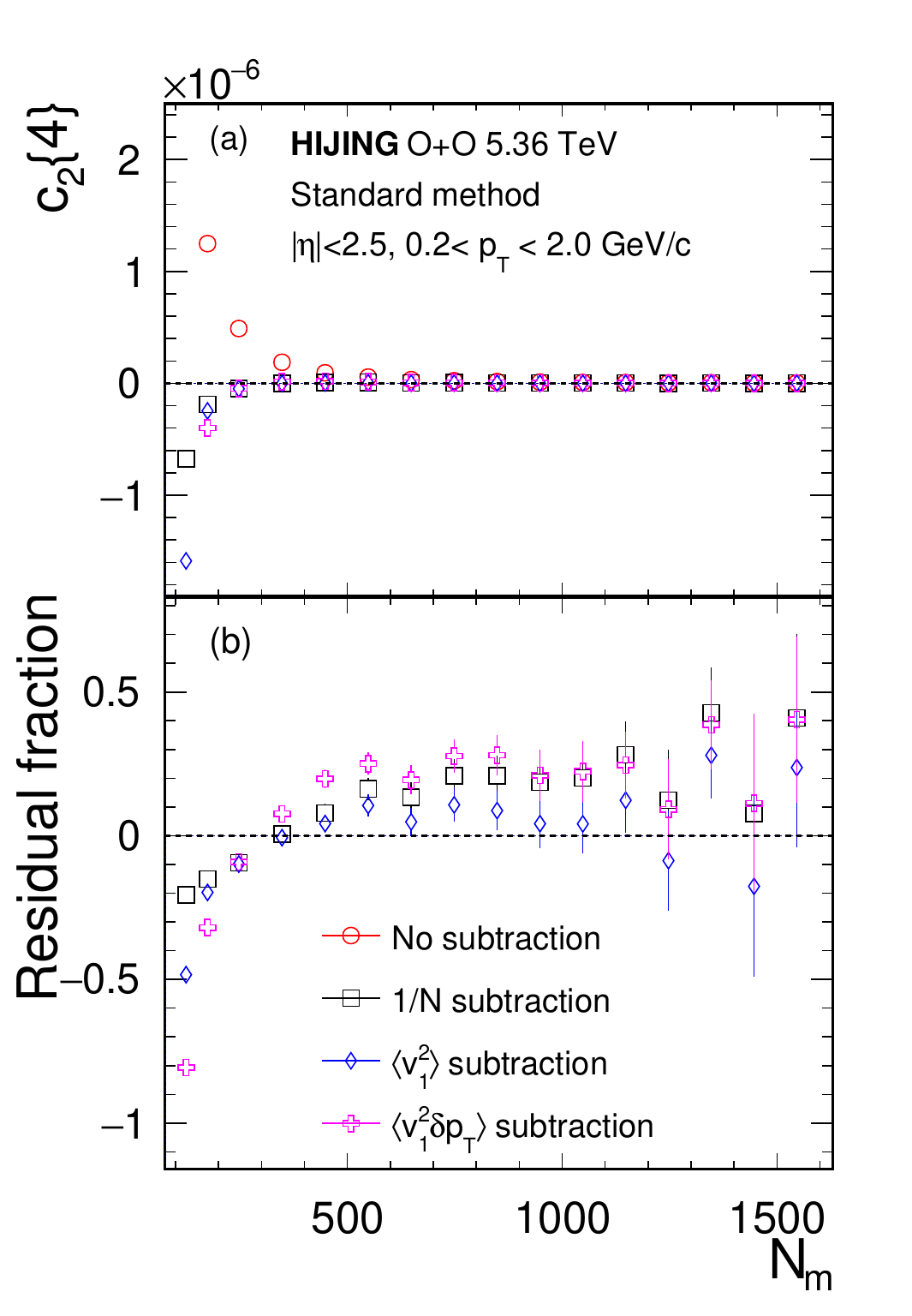}
\caption{\textbf{Nonflow subtraction for $c_2\{4\}$} for O+O collisions at 5.36 TeV in standard method as a function of $N_m$. The top panel shows the $c_2\{4\}$ before and after subtraction. The bottom panel shows the fractional residual nonflow defined in Eq.~\eqref{eq:4.12} using $1/N$-scaling (open boxes), $\lr{v_1^2}^{\obs}$-scaling (open diamonds), and $\lr{v_1^2\,\dpt}^{\obs}$-scaling (open crosses).}
\label{fig:4}
\end{figure}
Finally, we study the performance for $c_2\{4\}$. Nonflow is already strongly suppressed in four-particle cumulants, and in the two-subevent method the residual signal is too small for a statistically meaningful closure test with the present sample. The quantitative validation is therefore restricted to the standard method, which retains a much larger nonflow signal. This restriction is a limitation of the present validation and reflects the available sample size rather than a limitation of the subtraction method. The result is shown in Fig.~\ref{fig:4} for O+O at $5.36$~TeV. Outside the lowest-multiplicity bin, the properly scaled subtraction brings the residual close to zero in all three reference schemes, with a small but systematic undersubtraction. The residual is smallest, around $10$--$20\%$, for the $\lr{v_1^2}^{\obs}$-scaling method, and increases to about $20$--$30\%$ for the $1/N$ and $\lr{v_1^2\,\dpt}^{\obs}$ scalings. Closure deteriorates in the lowest-multiplicity bin. Two effects contribute to this deterioration. The source counting behind Eq.~\eqref{eq:4.1} requires many sources, and their number is smallest at low multiplicity, where departures from $1/N^{m-1}$ scaling are therefore largest. In addition, the subtracted term approaches the observed one at low $N$, so the residual becomes a small difference between two large numbers and any scaling mismatch is amplified in the residual fraction. The present study does not define a universal lower multiplicity limit. In data, that limit depends on the model non-closure, the magnitude of the genuine flow signal, and possible flow in the reference sample, all of which vary with system and observable. In particular, for the same non-closure fraction, a larger genuine flow signal relative to the residual nonflow would naturally permit the subtraction to be extended to lower multiplicity.

\begin{figure*}[tp]
\centering
\includegraphics[width=0.85\linewidth]{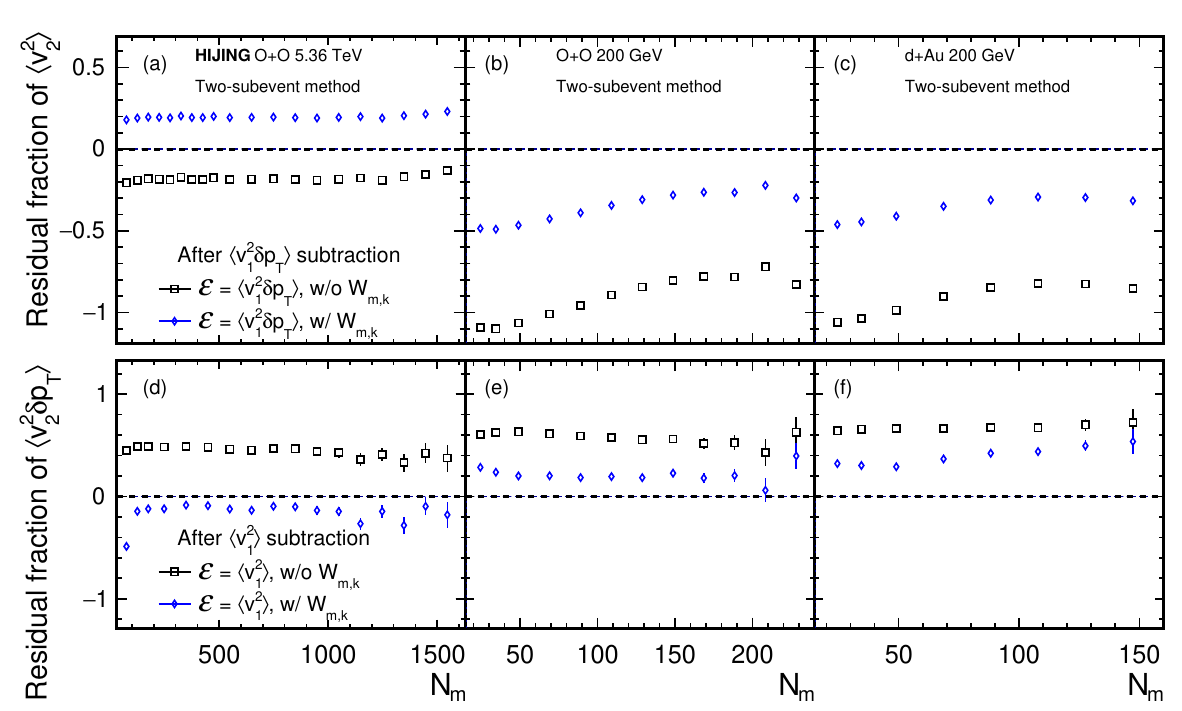}
\caption{\textbf{Effect of the multiplicity-reweighting factor with $v_1$-based nonflow references}. 
Impact of including the multiplicity-reweighting factor for two target-reference combinations: $\Ocal = \lr{v_2^2}$ and $\Ecal=\lr{v_1^2\,\dpt}^{\obs}$ (top row), and $\Ocal=\lr{v_2^2\,\dpt}$ and $\Ecal=\lr{v_1^2}^{\obs}$ (bottom row) in O+O collisions at 5.36 TeV (left), O+O collisions at 200 GeV (middle), and $d$+Au collisions at 200 GeV (right) in two-subevent method as a function of $N_m$. Black squares: naive subtraction without $W_{m,k}$. Blue diamonds: correctly scaled subtraction with $W_{m,k}$.}
\label{fig:5}
\end{figure*}
\begin{figure*}[tp]
\centering
\includegraphics[width=0.85\linewidth]{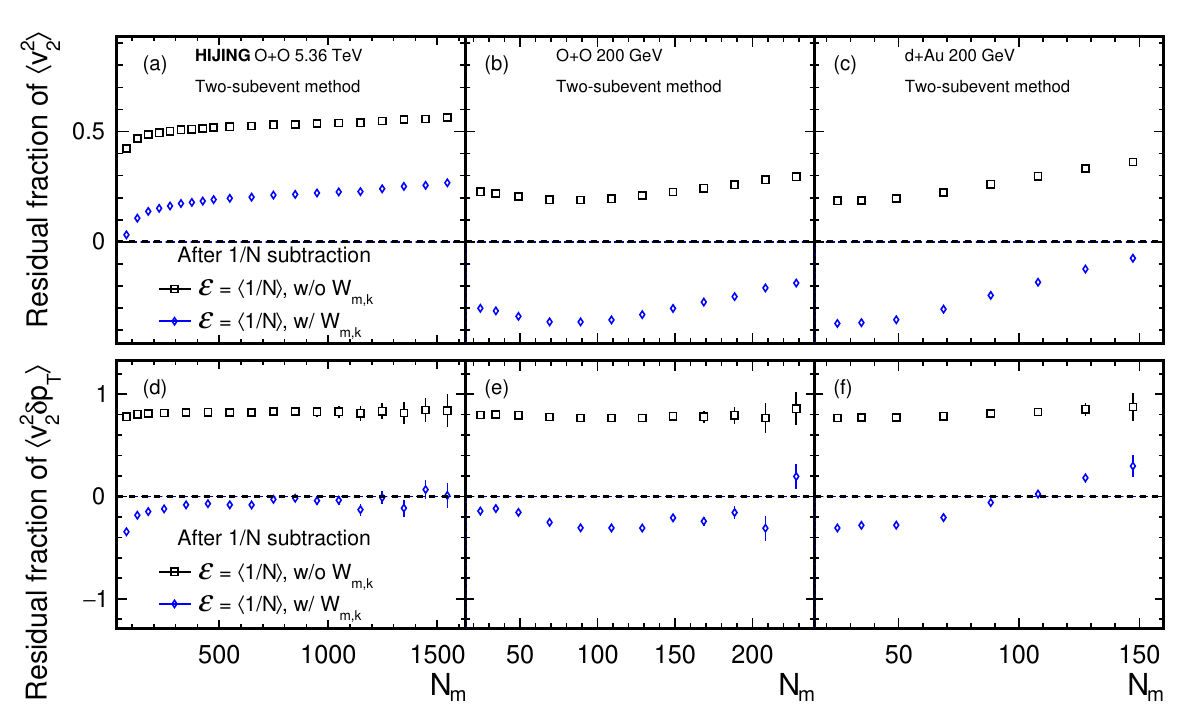}
\caption{\textbf{Effect of the multiplicity-reweighting factor with $1/N$-based nonflow reference}.  Impact of including the multiplicity-reweighting factor for two target-reference combinations: $\Ocal = \lr{v_2^2}$ and $\Ecal=1/N$ (top row)  and  $\Ocal=\lr{v_2^2\,\dpt}$ and $\Ecal=1/N$ (bottom row) in O+O collisions at 5.36 TeV (left), O+O collisions at 200 GeV (middle), and $d$+Au collisions at 200 GeV (right) in two-subevent method as a function of $N_m$. Black squares: naive subtraction without $W_{m,k}$. Blue diamonds: correctly scaled subtraction with $W_{m,k}$.}
\label{fig:6}
\end{figure*}
\subsection{Importance of the multiplicity-weighting} \label{sec:res3}
\begin{table}[t]
\centering
\caption{Effective multiplicities for the minimum-bias $pp$ reference samples. Here $N_1\equiv\lr{N}$, while $N_m$ for $m\geq2$ follows Eq.~\eqref{eq:4.2}. The two-subevent values use the corresponding subevent combination weights. They control the weight factor $W_{m,k}\approx N_{m,\pp}/N_{k,\pp}$ in Eq.~\eqref{eq:4.11}. The generator-level values use the charged-plus-neutral multiplicity definition specified in Sec.~\ref{sec:analysis}.}
\label{tab:2}
\begin{tabular}{c c c c c c}
\hline\hline
$\sqrt{s}$ & Method & $N_{1,\pp}$ & $N_{2,\pp}$ & $N_{3,\pp}$ & $N_{4,\pp}$ \\
\hline
5.36 TeV & Standard & 35.46 & 58.77 & 86.02 & 111.86 \\
5.36 TeV & Two-subevent & 35.46 & 59.48 & 87.44 & 113.66 \\
200 GeV & Two-subevent & 9.27 & 15.59 & 21.96 & 27.72 \\
\hline\hline
\end{tabular}
\end{table}

When the $\LM$ sample spans a wide multiplicity range, the multiplicity-reweighting factor $W_{m,k}$ in Eq.~\eqref{eq:4.10} is required within the independent-source construction. The corresponding effective multiplicities for the $pp$ reference are summarized in Table~\ref{tab:2}. They increase significantly with the order of the correlator, by about a factor of three from $N_{1,\pp}$ to $N_{4,\pp}$ in every sample, because a higher-order correlator weights the broad $pp$ multiplicity distribution more heavily towards its high-$N$ tail. This is in sharp contrast with the target observable, where the narrow multiplicity bins mean that $N_k$ barely changes with $k$ and $N_k/N_m\simeq1$, so $W_{m,k}$ is dominated by the constant reference ratio $N_{m,\pp}/N_{k,\pp}$. We illustrate this for the three target observables, showing that neglecting $W_{m,k}$ when using minimum-bias $pp$ (or a very wide peripheral bin) as the reference leads to a subtraction failure whose severity grows with the correlator order. The residual dependence of $W_{m,k}$ on the target multiplicity is negligible on the scale of Fig.~\ref{fig:5}.

The top panels of Fig.~\ref{fig:5} show the residual nonflow fraction for $\lr{v_2^2}$ with $\lr{v_1^2\,\dpt}^{\obs}$ as the reference. Since $W_{2,3} \approx N_{2,\pp}/N_{3,\pp} < 1$, neglecting $W$ always oversubtracts. At LHC O+O, including $W$ instead flips the residual to the undersubtracted side with similar magnitude. Here the $\lr{v_1^2\,\dpt}^{\obs}$ reference simply does not capture the $N$-dependence of the $\lr{v_2^2}$ nonflow, regardless of $W$. The $W$ correction is nevertheless required: it removes a well-defined $pp$-weighting bias and is needed for consistency across systems. At RHIC, in both O+O and $d$+Au, including $W$ improves the subtraction substantially and pulls the residual fraction much closer to zero. 

The bottom panels of Fig.~\ref{fig:5} show the residual nonflow fraction for $\lr{v_2^2\,\dpt}$ using $\lr{v_1^2}^{\obs}$ as the reference observable. Here $W_{3,2} \approx N_{3,\pp}/N_{2,\pp} > 1$, so neglecting $W$ always leads to undersubtraction. In all three collision systems, including $W$ improves the subtraction substantially and brings the residual nonflow fraction much closer to zero. The improvement is more dramatic here than in the top panels because $W_{3,2}$ enters squared in the bottom-panel case, whereas $W_{2,3}$ enters to the first power in the top-panel case.

Figure~\ref{fig:6} shows the analogous comparison in the $1/N$-scaling method when $N_{m,\pp}$ is replaced by the simple event-averaged multiplicity $\lr{N}_{\mathrm{pp}}$ in $pp$, which is always smaller. As expected, omitting the reweighting substantially underestimates the subtraction. For $\Ocal = \lr{v_2^2}$ with $\Ecal=1/N$ the scale factor is suppressed by a factor of $\lr{N}_\pp/N_{2,\pp}$, and for $\Ocal=\lr{v_2^2\,\dpt}$ with $\Ecal=1/N$ it is suppressed by $(\lr{N}_\pp/N_{3,\pp})^2$, a much larger reduction. The naive $1/N^{m-1}$ scaling without the proper $pp$ reweighting therefore significantly underestimates the nonflow for $\lr{v_2^2}$, with an even more severe effect for $\lr{v_2^2\,\dpt}$. As shown in the bottom panels of Fig.~\ref{fig:6}, the residual fraction for $\lr{v_2^2\,\dpt}$ remains near unity, that is, a nearly complete undersubtraction in which the naive scale factor is so small that almost none of the observed nonflow is removed; with the proper reweighting it is driven to zero within statistical uncertainties outside the lowest-multiplicity region.

\begin{figure}[htb]
\centering
\includegraphics[width=0.8\linewidth]{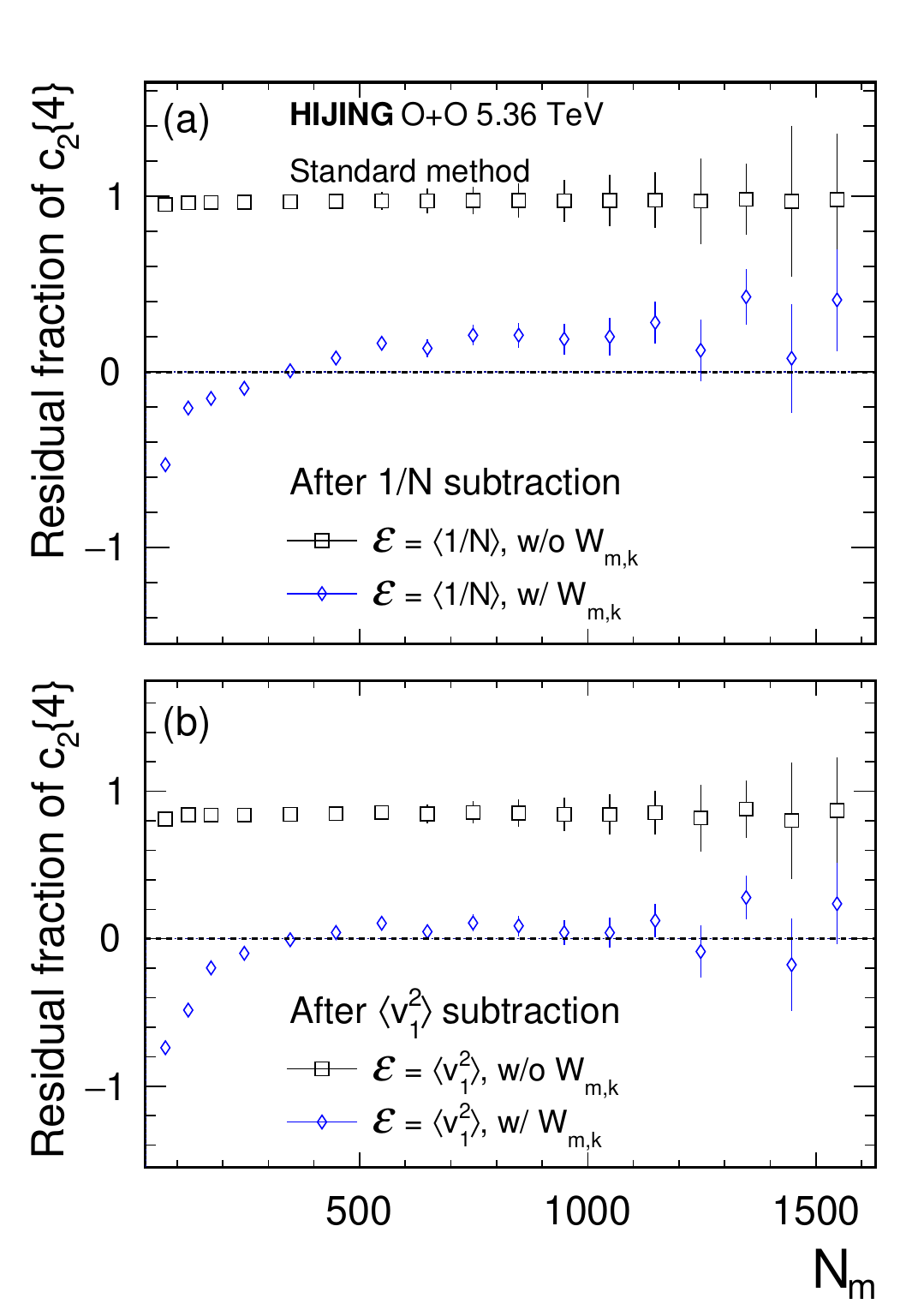}
\caption{\textbf{Effect of the multiplicity-reweighting factor for $c_2\{4\}$}. Impact of including the multiplicity-reweighting factor for reference observable $\Ecal = 1/N$ (top) and $\Ecal=\lr{v_1^2}^{\obs}$ (bottom) in O+O collisions at 5.36 TeV in standard method as a function of $N_m$. Black squares: naive subtraction without $W_{m,k}$. Blue diamonds: correctly scaled subtraction with $W_{m,k}$.}
\label{fig:7}
\end{figure}
We repeat the same study for the four-particle cumulant $c_2\{4\}$ from the standard method with two reference observables, $\Ecal = \lr{v_1^2}^{\obs}$ and $\Ecal = 1/N$. The results are shown in Fig.~\ref{fig:7}. Because the reweighting factor enters to the \emph{third power}, the impact of omitting it is much more severe. For illustration, an approximately $20\%$ upward deviation of $W$ from unity produces an approximately $70\%$ change in $W^3$. The factors that apply in Fig.~\ref{fig:7} are $W_{4,1}$ for the $1/N$ reference and $W_{4,2}$ for the $\lr{v_1^2}^{\obs}$ reference; both are substantially larger than unity (Table~\ref{tab:2}). Without the reweighting factor, the residual fraction remains near unity. Including it improves closure over most of the multiplicity range, but the lowest-multiplicity point remains poorly described and can be oversubtracted.

The same effect also resolves a long-standing observation in 2PC: the $c_0$ method is known to undersubtract nonflow in data~\cite{STAR:2023wmd,ATLAS:2015hzw}. Within the present framework the explanation is direct. The correctly scaled $c_0$ subtraction for $\lr{v_2^2}$ reads
\begin{equation}
  \frac{\lr{v_2^2}}{\lr{v_2^2}^{\obs}}
  = 1 - \frac{N_{2,\pp}}{N_2}
    \frac{\lr{v_2^2}_\pp^{\obs}}{\lr{v_2^2}^{\obs}}.
  \label{eq:6.1}
\end{equation}
The numerator requires $N_{2,\pp}$, the effective pair multiplicity in the $pp$ reference sample, not the mean $\lr{N}_\pp$. Because the $pp$ distribution is broad, $N_{2,\pp} = \lr{N^2}_\pp/\lr{N}_\pp > \lr{N}_\pp$. The naive $c_0$ implementation uses $\lr{N}_\pp$ instead of $N_{2,\pp}$, causing systematic under-subtraction. 

\subsection{Model non-closure and its propagation to data}
\label{sec:closure}

Even with the optimal subtraction, the residual fraction is not exactly zero, reflecting deviations of the actual nonflow from perfect $1/N^{m-1}$ scaling in \HIJING{}. The non-closure is at the 20--30\% level for most two-particle correlation observables but can be significantly larger for higher-order correlations.

This non-closure cannot be corrected to zero in data without knowing the true nonflow. We therefore define a multiplicative correction factor $K_{\Ocal}^{\Ecal}(N)$ that quantifies the fractional non-closure in models:
\begin{equation}
  K_{\Ocal}^{\Ecal}(N)
  \equiv \left(\frac{R_\Ocal}{R_\Ecal}\right)^{m-1}
  \;\approx\; 1
   + \frac{\Ocal\{m\}^{\mathrm{res,\,nonflow}}}{\Ocal\{m\}^{\obs}}\,,
  \label{eq:6.2}
\end{equation}
Therefore, for a small residual fraction, $K_{\Ocal}^{\Ecal}-1$ approximately equals the residual fractional nonflow shown
in Figs.~\ref{fig:2}--\ref{fig:7}.

By construction, the subtraction formula Eq.~\eqref{eq:4.12} closes in \HIJING{} when the scale factor is rescaled by $K_{\Ocal}^{\Ecal}$:
\begin{equation}
  0 \;=\; 1 - K_{\Ocal}^{\Ecal}
       \!\left(\frac{R_\Ecal}{R_\Ocal}\right)^{m-1}\!.
  \label{eq:6.3}
\end{equation}

The absolute nonflow magnitude in \HIJING{} is not reliable, but predicting the trend of the nonflow is a more robust exercise than predicting its magnitude, and $K_{\Ocal}^{\Ecal}(N)$ is a measure of that trend. This follows from its definition: $K_{\Ocal}^{\Ecal}$ is the double ratio $(R_\Ocal/R_\Ecal)^{m-1}$, in which the overall normalization of the nonflow, the quantity \HIJING{} gets wrong, cancels between target and estimator measured in the same events, leaving only the mismatch in how the two scale with $N$. We therefore propose that this fractional trend can be more transferable than the absolute \HIJING{} nonflow yield. Under the explicit assumption that the same fractional non-closure applies in data, $K_{\Ocal}^{\Ecal}(N)$ can be used as a multiplicative correction to the nonflow baseline measured in peripheral or $pp$ data; the absolute normalization of that baseline remains data driven. This is a model-transfer assumption, and its uncertainty should be constrained by comparisons among generators and by data-driven consistency tests. The STAR 2PC procedure~\cite{STAR:2023wmd} provides a valuable precedent for using \HIJING{} to assess a subtraction uncertainty rather than to predict the absolute nonflow yield.

The discussion also suggests how the estimator should be chosen in data. Section~\ref{sec:res2} shows that the best estimator depends on the target observable and on the collision energy, so the ranking found in \HIJING{} should not simply be carried over to a measurement. What we recommend instead is to carry all the available estimators through the analysis for the system and energy at hand, to quote as the central value the one best motivated for the target observable, and to take the spread among them, together with the deviation of $K_{\Ocal}^{\Ecal}$ from unity, as the systematic uncertainty of the subtraction. This is data driven and does not require the \HIJING{} ranking to be right. A large spread is itself the signal that no single scaling describes the nonflow in that system, and that the subtracted signal should be read with care.

\section{Summary and outlook}
\label{sec:summary}

Multi-particle cumulants are among the most powerful probes of collective behavior, but in small collision systems their interpretation is limited by residual nonflow that survives subevent suppression. We have developed a general subtraction framework for an $m$-particle cumulant and tested its implementation in \HIJING{} for two-, three-, and four-particle observables. The analysis identifies $v_1$-containing correlators as nonflow-dominated estimators and demonstrates the importance of the multiplicity-reweighting correction, while also quantifying the model non-closure and its deterioration in the lowest-multiplicity region. The main findings are:

\begin{itemize}
\item Nonflow in an $m$-particle correlator scales approximately as $1/N^{m-1}$ in the independent-source picture (Eq.~\ref{eq:4.1}). This scaling has been largely confirmed in the \HIJING{} model. 

\item Correlators containing $v_1$ are nonflow-dominated estimators at each order [Eqs.~\eqref{eq:4.4}--\eqref{eq:4.9}], because the $\pT$-integrated genuine dipolar flow is expected to be small, while the surviving low-$\pT$ long-range nonflow includes a GMC-induced dipolar component with the expected $1/N^{m-1}$ scaling.

\item For 2PC the $1/N$-scaling subtraction leaves a residual nonflow fraction below 20--30\%; for multi-particle correlations the residual can be larger. Choosing an estimator that follows the target nonflow more closely can reduce these residuals further.

\item Within the independent-source construction, the multiplicity-reweighting factor $W_{m,k}$ associated with a broad $pp$ or peripheral reference is required in Eq.~\eqref{eq:4.10}. Its omission produces errors amplified by the power $m-1$: for $c_2\{4\}$, the residual nonflow fraction stays near 1 without $W_{m,k}$ but is much closer to zero with it outside the lowest-multiplicity region. This factor also explains the well-known under-subtraction of the naive $c_0$ method in 2PC.

\item The optimal scaling estimator depends on the target observable. Of the three tested here, $c_0$ ($1/N$), $c_1$ ($\lr{v_1^2}^{\obs}$), and $\lr{v_1^2\,\dpt}^{\obs}$, the $c_1$ scaling is best for $\lr{v_2^2}$ ($\lesssim 15\%$ residual) and for $c_2\{4\}$ ($\sim 10$--$20\%$), while $c_0$ is best for $\lr{v_2^2\,\dpt}$ ($< 20\%$). The remaining combinations leave $\sim 20$--$50\%$ residuals. These ranges refer to the intermediate- and high-multiplicity region, and the $c_2\{4\}$ result is validated only for the standard method. The ranking is specific to \HIJING{}; in data, all viable estimators should be carried through the analysis and their spread treated as part of the subtraction uncertainty, as discussed in Sec.~\ref{sec:closure}.

\item The residual fractional non-closure in \HIJING{} is quantified by $K_{\Ocal}^{\Ecal}(N)$ in Eq.~\eqref{eq:6.2}. Under the stated assumption that this fraction is transferable, it can be propagated to the data-driven nonflow baseline as a model-based non-closure uncertainty. Cross-generator and data-driven consistency tests are needed to constrain this assumption.
\end{itemize}

Several directions merit future work.
\begin{itemize}

\item \textit{Cross-generator and flow-plus-nonflow validation.} The $1/N^{m-1}$ scaling follows from the independent-source picture, but the size of its violation, the estimator ranking, and $K_{\Ocal}^{\Ecal}(N)$ do not. Repeating the closure test with generators having different nonflow composition, and with a generator or afterburner carrying both genuine flow and nonflow, would test the recovery of a known flow signal.
%which the present nonflow-only study cannot do.

\item \textit{Symmetric cumulants and mixed-harmonic correlators.} The framework extends naturally to $\mathrm{sc}(n,m)$, $\lr{V_2^2 V_4^*}$, and $\lr{V_2 V_3 V_5^*}$ via Method 2b, with the subtraction performance to be validated in \HIJING{} and \PYTHIA{}.

\item \textit{Higher-order harmonics and cumulants.} The framework presented here focuses on $n=2$. Extension to $v_3$- and $v_4$-related observables, where the dominant nonflow source may be different, requires further dedicated study. Natural candidates for $v_3$ are $C_{112}$ and $C_{123}$ [Eqs.~\eqref{eq:4.7}--\eqref{eq:4.8}]~\cite{STAR:2017idk}. Methods 2a and 2b also generalize naturally to higher-order cumulants $c_n\{6\}$ and $c_n\{8\}$, where the multiplicity-reweighting correction becomes increasingly important as the correlator order grows.

\item \textit{Generalization of the template method.} A multi-particle version of the template method outlined in Eqs.~\eqref{eq:4.14} and \eqref{eq:4.15} would account for the presence of genuine flow in $pp$, and hence avoid the subtraction bias in the low-$N$ region. 

\item \textit{Consistency between standard and subevent methods.} These methods carry the same genuine flow signal but very different nonflow contamination. The degree of agreement after subtraction therefore provides a data-driven consistency test.
%, while common biases must be assessed separately.

\item \textit{Application to data.} The most important next step is to apply these methods in real RHIC and LHC analyses. ATLAS, CMS, ALICE, STAR, and sPHENIX are all positioned to implement the framework on existing $pp$, $p$+Pb/Au, $\dAu$, and $\OO$ datasets. Under the stated transfer assumption, the $K_{\Ocal}^{\Ecal}$ correction provides a model-based non-closure uncertainty whose range can be constrained through generator comparisons and consistency between standard and subevent methods. The resulting improvements in multi-particle flow measurements should substantially sharpen the picture of small-system collectivity.
\end{itemize}

{\bf Acknowledgments}  We thank J. Grosse-Oetringhaus for valuable comments. Z. Wang and J. Chen are supported by the National Key Research and Development Program of China under Contract No. 2022YFA1604900 and the National Natural Science Foundation of China (NSFC) under Contract No. 12625501. J. Jia, Z. Yan, and S. Huang are supported by the U.S. Department of Energy, Office of Science, Office of Nuclear Physics, under DOE Awards No. DE-SC0024602. C. Zhang is supported by the National Key Research and Development Program of China under Contract No. 2024YFA1612600, Shanghai Pujiang Talents Program under Contract No. 24PJA009 and the NSFC under Contract No. 12547102.

{\bf Author contributions} J.\ Jia conceived the study. Z.\ Wang carried out the data analysis under the supervision of J.\ Chen, C.\ Zhang, and J.\ Jia. J.\ Jia and S.\ Huang led the writing of the manuscript. All authors discussed the results and edited the final version. Claude and ChatGPT AIs were used to verify formulas and to improve the paper draft.

%\clearpage
\appendix

\section{The effective multiplet multiplicity $N_m$}
\label{app:Nn}

For an event with $N$ particles, the number of $m$-particle combinations is $P_m(N) = N(N-1)\cdots(N-m+1)$. For a sample of events with multiplicity distribution $p(N)$, the sample-averaged correlator is weighted by these combination counts:
\begin{equation}
  \Ocal\{m\}
  = \frac{\sum_i P_m(N_i)\, \Ocal\{m\}(N_i)}
         {\sum_i P_m(N_i)}.
  \label{eq:A.1}
\end{equation}
The effective multiplicity $N_m$ used throughout this paper is defined by averaging $1/N$ over the same combinations,
\begin{equation}
  \frac{1}{N_m}
  \equiv \frac{\sum_i P_m(N_i)/N_i}
         {\sum_i P_m(N_i)},
  \label{eq:A.2}
\end{equation}
which is Eq.~(\ref{eq:4.2}). The averaged quantity, $1/N$, is the same for every order; only the weight $P_m(N)$ depends on $m$. Using $P_m(N_i) \approx N_i^m$ for $N_i \gg m$ gives $P_m(N_i)/N_i \approx N_i^{m-1}$, so
\begin{equation}
  \frac{1}{N_m} \approx \frac{\sum_i N_i^{m-1}}{\sum_i N_i^m},
  \quad\Longrightarrow\quad
  N_m \approx \frac{\sum_i N_i^m}{\sum_i N_i^{m-1}} = \frac{\lr{N^m}}{\lr{N^{m-1}}},
  \label{eq:A.3}
\end{equation}
which is Eq.~(\ref{eq:4.3}). The corrections are suppressed by inverse powers of $N$. In this form $N_m$ is the ratio of consecutive moments of the multiplicity distribution.  It increases with $m$ and satisfies $N_m\geq N_1=\lr{N}$. This definition is adopted for the reasons given in Sec.~\ref{sec:Nm}.

For $m=2$, $N_2 \approx \lr{N^2}/\lr{N} = \lr{N} + \sigma_N^2/\lr{N}$. A broad $pp$ multiplicity distribution can therefore give $N_2$ well above $\lr{N}$. For higher orders, the stronger weight of high-multiplicity events increases the differences among the effective multiplicities. For example, Table~\ref{tab:2} gives $N_{4,\pp}/N_{2,\pp}$ of order $1.5$--$2$. This ratio enters to the third power when a two-particle estimator is used for a four-particle target, so its effect is of order $3$--$8$. This order-dependent weighting is the correction missed when the effective-multiplicity ratio is omitted.

For the subevent method, the combination weight in Eq.~(\ref{eq:A.2}) is modified. For the two-subevent three-particle correlator, the weight is $N_a N_b (N_a + N_b - 2)$, and $N_m$ must be recalculated accordingly. The conceptual definition in Eq.~(\ref{eq:A.2}) and its physical interpretation remain unchanged.

\section{The alternative effective multiplicity $M_m$}
\label{app:Mn}

If nonflow scaled exactly as $\Ocal\{m\}^{\nf} = C/N^{m-1}$ for some constant $C$, the sample-averaged nonflow from Eq.~\eqref{eq:A.1} would be
\begin{equation}
  \Ocal\{m\}^{\nf}
  = C \cdot
    \frac{\sum_i P_m(N_i)/N_i^{m-1}}
         {\sum_i P_m(N_i)}.
  \label{eq:B.1}
\end{equation}
This suggests an alternative effective multiplicity $M_m$, defined via $\Ocal\{m\}^{\nf} \equiv C/M_m^{m-1}$, i.e., from the combination-weighted average of $1/N^{m-1}$ rather than of $1/N$:
\begin{equation}
  \frac{1}{M_m^{m-1}}
  \equiv \frac{\sum_i P_m(N_i)/N_i^{m-1}}
         {\sum_i P_m(N_i)}.
  \label{eq:B.2}
\end{equation}
$M_m$ is the fixed multiplicity that reproduces the same combination-weighted nonflow when the $1/N^{m-1}$ scaling is exact. Using $P_m(N_i)/N_i^{m-1} \approx N_i$ for $N_i \gg m$ gives
\begin{equation}
  M_m^{m-1} \approx \frac{\sum_i N_i^m}{\sum_i N_i},
  \label{eq:B.3}
\end{equation}
to be compared with $N_m\approx\sum_iN_i^m/\sum_iN_i^{m-1}$ in Eq.~\eqref{eq:A.3}.

For $m=2$ the two definitions coincide. For $m\geq3$, the approximate forms give
\begin{align}
  M_3^2 \approx N_3 \cdot N_2,\;  M_4^3 \approx N_4 \cdot N_3 \cdot N_2,\;\ldots
  \label{eq:B.4}
\end{align}
The convexity of $x^{m-1}$ for $x=1/N>0$ also gives $M_m\leq N_m$.

To express these relations at non-integer order, we extend the moment-ratio approximation by defining $N_x\equiv\lr{N^x}/\lr{N^{x-1}}$ for real $x$, using the same event average for all orders. If $\ln N_x$ varies approximately linearly over the relevant range of $x$, the geometric mean in Eq.~\eqref{eq:B.4} can be approximated by its midpoint value, $M_m\approx N_{(m+2)/2}$, giving
\begin{equation}
  M_3 \approx N_{2.5},\quad M_4 \approx N_3,\quad M_5 \approx N_{3.5},
  \quad \ldots
  \label{eq:B.5}
\end{equation}

For the narrow target bins, $M_m\simeq N_m\simeq\lr{N}$. Within this approximation, the distinction between the two definitions enters only through the broad $pp$ reference, where the numerical results use $N_m$ in Eq.~\eqref{eq:4.10}. Checks with $M_m$ show that $N_m$ gives better closure in many of the mixed-order cases studied in \HIJING{}. The difference matters most at higher order because the $pp$ effective-multiplicity ratio enters with a power in the subtraction.

We do not have a full explanation of why $N_m$ outperforms $M_m$, despite that $M_m$ seems more natural. One possible interpretation is that the nonflow may fall more slowly than $1/N^{m-1}$ within the broad $pp$ sample, or the multiplicity may be correlated with jet activity. The preference may also be specific to \HIJING{} model and might be very different in the real data. We therefore treat the difference between $N_m$ and $M_m$ as model dependence and leave a fuller study to future work.

\section{Subtraction with a peripheral-bin reference}\label{app:peri}
\begin{figure}[htb]
\centering
\includegraphics[width=0.9\linewidth]{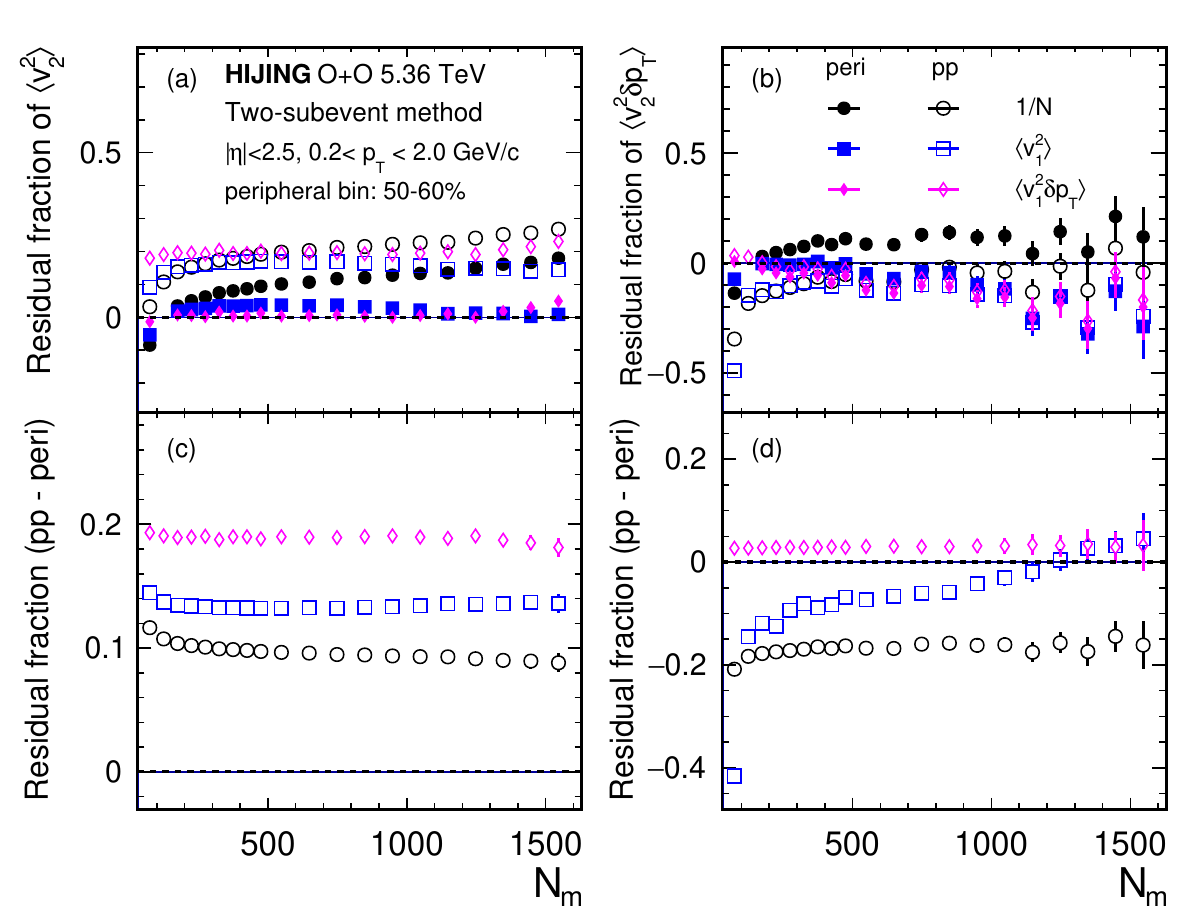}
\caption{\textbf{Comparison of minimum-bias $pp$ and peripheral-bin references.} Residual nonflow fractions for $\lr{v_2^2}$ (a) and $\lr{v_2^2\,\dpt}$ (b) in O+O collisions at $\snn=5.36$~TeV using the two-subevent method. Filled symbols use the 50--60\% O+O peripheral bin as the low-multiplicity reference, while open symbols use minimum-bias $pp$ collisions. Different symbols correspond to the three nonflow-scaling prescriptions considered in the main text. Panels (c) and (d) show the difference between the residual fractions obtained with the $pp$ and peripheral-bin references for the corresponding observables.}
\label{fig:9}
\end{figure}

Peripheral-bin subtraction has often been used in two-particle correlation analyses, as discussed in Sec.~\ref{sec:2pc}. The present framework can be applied to such a reference in the same way as to minimum-bias $pp$. A peripheral bin is affected by event-selection bias, so the non-closure performance depends on the chosen interval. Figure~\ref{fig:9} uses a 50--60\% O+O peripheral interval as a concrete test and compares it with the minimum-bias $pp$ reference for O+O collisions at $\sqrt{s_{NN}}=5.36$~TeV. The effective multiplicities and scale factors for each reference are calculated with the same order-dependent combination weights used for the corresponding target and estimator observables. The residual nonflow fractions are generally less than 20\% over most of the multiplicity range, better than the subtraction based on the $pp$ reference. However, the suitability of the 50--60\% interval is model and system dependent. This reference dependence is one reason that minimum-bias $pp$ is retained as the default reference in the main analysis.

\bibliography{ref}{}
\bibliographystyle{apsrev4-1}
\end{document}